\documentclass[letterpaper,journal,final]{IEEEtran}  

\IEEEoverridecommandlockouts                              

\usepackage{graphicx,psfrag}
\usepackage{amsmath,amssymb,amsfonts,mathrsfs,mathtools}
\usepackage{color,epstopdf}
\usepackage{algorithmic}
\usepackage{enumerate} 
\usepackage{subfigure}
\usepackage{floatrow}
\usepackage{float}
\usepackage{stfloats}
\usepackage{lipsum}
\usepackage{flushend}
\usepackage{bbold}
\usepackage{dsfont}
\usepackage{mathrsfs}
\usepackage{relsize}
\usepackage{relsize}
\usepackage[normalem]{ulem}
\usepackage{url}

\newtheorem{example}{Example}

\newtheorem{assumption}{Assumption}

\newtheorem{remark}{Remark}
\newtheorem{theorem}{Theorem}

\newtheorem{lem}{Lemma}

\newtheorem{proposition}{Proposition}

\newcommand{\bbm}{\begin{bmatrix}}
\newcommand{\ebm}{\end{bmatrix}}

\newcommand\oprocendsymbol{\hbox{$\square$}}
\newcommand\oprocend{\relax\ifmmode\else\unskip\hfill\fi\oprocendsymbol}

\def\qedp{\hspace*{\fill}~{\tiny $\blacksquare$}}
\def\qeds{\hspace*{\fill}~{\tiny $\square$}}
\def\be{\begin{equation}}
\def\ee{\end{equation}}
\def\ba{\begin{array}}
\def\ea{\end{array}}
\def\eqa{\begin{eqnarray}}
\def\eqe{\end{eqnarray}}

\definecolor{darkgreen}{rgb}{0.0, 0.55, 0.0}
\definecolor{amaranth}{rgb}{0.9, 0.17, 0.31}
\usepackage[prependcaption,colorinlistoftodos]{todonotes}

\title{\LARGE \bf Feedback linearization through the lens of data}

\author{C. De Persis, D. Gadginmath, F. Pasqualetti and P. Tesi
\thanks{This material is based upon work supported
    in part by the project Digital Twin with project number P18-03 of
the research programme TTW Perspective which is (partly)
financed by the Dutch Research Council (NWO) and by awards ARO W911NF-20-2-0267, AFOSR-FA9550-20-1-0140,
    AFOSR-FA9550-19-1-0235, and NSF-2020246.
    C. De Persis is with ENTEG, University of Groningen, 9747AG Groningen, 
The Netherlands. Email: {\tt\small c.de.persis@rug.nl}. 
D. Gadginmath and F. Pasqualetti are with the 
Department of Mechanical Engineering, University of California at Riverside, Riverside, CA, 92521, USA. Email: {\tt\small dgadg001,fabiopas@engr.ucr.edu}. 
P. Tesi is with the Department of Information Engineering, 
University of Florence, 50139 Florence, Italy. Email: {\tt\small pietro.tesi@unifi.it}. 
}
}

\begin{document}

\maketitle


\begin{abstract}
Controlling nonlinear systems, especially when data are being used to offset uncertainties in the model, is hard. A natural approach when dealing with the challenges of nonlinear control is to reduce the system to a linear one via change of coordinates and feedback, an approach commonly known as feedback linearization. Here we consider the feedback linearization problem of an unknown system when the solution must be found using experimental data. We propose  
a new method that learns the change of coordinates and the linearizing controller from a library (a dictionary) of
candidate functions with a simple algebraic procedure -- the computation of the null space of a data-dependent matrix. Remarkably, we show that the solution is valid over the 
entire state space of interest and not just on the dataset used to determine the solution.
\end{abstract}


\section{Introduction}

Data-driven control is becoming more and more central to 
control engineering. The motivation is to automate the control design process 
when the system of interest is poorly modeled (first-principles laws might be difficult 
to obtain) or when accurate models are too complex to be used for controller design. 
Data-driven control has been applied in almost every area
of control theory, among which we can mention optimal 
\cite{baggio2019data,dean2019sample,de2019formulas,Pellegrinoi2022ECC}, 
robust \cite{de2019formulas,berberich2020robust,henk-ddctr-uncer,BisoffiAUT2021Petersen},
and predictive control \cite{deepc,berbMPC2020,lian2021nonlinear,coulson2021distributionally}. 
Further applications include networked and distributed control 
\cite{baggio2021net,Celi2021CDC,allibhoy2020data,Sforni2022CDC}. Also the techniques are quite varied,
ranging from dynamical programming and behavioral theory to techniques typical of machine learning. 
The majority of the results considers linear systems.
Unsurprisingly, deriving solutions for {nonlinear} systems is
harder. The objective of this paper is to explore data-driven control for nonlinear systems,
in particular data-driven \emph{feedback linearization}. 

\emph{Related work.} Data-driven control for nonlinear systems is still largely unexplored.
A way to deal with nonlinear systems is to exploit some structure, when it is a priori known the class to which the
system belongs, for instance bilinear \cite{bisoffi2020data,yuan2021data}, and
polynomial (or rational) systems \cite{BisoffiAUT2021Petersen,guoTAC2021poly,dai2020semi,Nejati2022poly,strasser2021data}.
Methods for general nonlinear systems include linearly parametrized models 
with basis functions \cite{dai2021statedependent,Chang2020basis,NonlinearityCancellation2023},
Gaussian process models \cite{Umlauft2018GP,Umlauft2020FL},
linear \cite{de2019formulas,
Fraile2020FL,luppi2022data,cheah2022robust} and polynomial approximations \cite{Guo2022Taylor,Martin2022Taylor}, and the
feedback linearization technique \cite{NonlinearityCancellation2023,Umlauft2020FL,Westenbroek2020FL,Darshan2022}, 
which is the focus of this work.

As it is well known, the feedback linearization method aims at finding a 
coordinate transformation where the dynamics can be linearized via feedback 
(i.e., where all the nonlinearities can be canceled out by a feedback controller) \cite{brockett1978feedback,Jakubczyk1980linearization}.
Linearization is clearly appealing as it allows us to exploit all the concepts and tools that are available 
for linear systems, thus simplifying problems that would otherwise be difficult to address.
In the context of data-driven control, recent examples
are sampled-data stabilization \cite{Fraile2020FL} and output-matching control \cite{alsalti2021data}.
We remark that also the problem of lifting a nonlinear system to a linear system  
of a \emph{higher} dimension has been studied, notable
methods are \emph{immersion} \cite{jungers2020immersion} 
and the \emph{Koopman operator theory} \cite{Kaiser2021,cortes2022koopman}.
In this respect, feedback linearization can be viewed as the natural starting point for techniques of this kind
because it searches for a mapping that \emph{preserves} the state dimension.
We refer the interested reader to \cite{Darshan2022} for  a recent discussion on the connections between 
feedback linearization and  Koopman operator theory.

\emph{Contribution.} In this paper we address the problem of feedback
linearization when the model is unknown yet data describing the system
dynamics is available. We focus on the simplified setting of
full-state linearization and noise-free data, as it allows us to
explain the intricacies of the problem while avoiding tangential
complexities. In fact, to the best of our knowledge, even this
simplified setting has remained unsolved, as most existing work
assumes the knowledge of the state coordinates that render the system
linearizable (we do not make such an assumption)
\cite{NonlinearityCancellation2023,Umlauft2020FL,Fraile2020FL,Westenbroek2020FL,alsalti2021data}.

We propose a data-driven feedback linearization approach where the
state and control transformations are learned as a combination of a
\emph{library} of basis functions. Approaches of this type have been
widely used in the context of system identification, see for example
\cite{Brunton2016}, and recently also in the context of direct
data-driven control
\cite{dai2021statedependent,NonlinearityCancellation2023,alsalti2021data}. We
start by revisiting the problem of model-based feedback linearization
(Section \ref{sec:MB}), which provides useful insights into the
required transformations and inspires our novel data-driven approach
(Section \ref{sec:DD}). Our data-driven feedback linearization
technique requires only a finite set of data points, yet it remains
provably applicable on the \emph{whole} space of interest. Further,
our solution features favorable computational properties, as it
requires only the computation of the null space of an appropriately
defined data matrix. Finally, although our approach makes use of the
derivatives of the state to compute the required data matrices, which
may result in numerical instabilities, we remark that alternative
approaches exist as we show, e.g., in \cite[Appendix
A]{cdp-rp-pt2023event}.
Preliminary results were  reported in our previous paper \cite{de2023data} and a few parts of this manuscript, including the problem setting and the model-based solution,  are taken from that source. However, this paper provides a different,  more general condition (Theorem \ref{thm:DD-2}) for the data-based solution of the problem than the one proposed in \cite{de2023data}. In fact, while in \cite{de2023data} the solution to the problem was found in the one-dimensional null space of a suitable matrix of data, here we show that, under a condition on the richness of data (Assumption \ref{asspt:data-rich}), 
we can allow the dimension of the null space to be larger than one. 
This is an important generalization because, as we discuss in Section \ref{ref:discussion}, by enlarging the dictionary of functions, possibly to raise our chances to have a complete dictionary, we might increase the nullity of the matrix of data used to determine a solution to the feedback linearization problem. In addition to the theoretical findings, 
new detailed numerical examples that illustrate the results have been added.

\emph{Notation and definitions.} Given a function $h:\mathcal{D} \to \mathcal{E}$,
its inverse function $h^{-1}:\mathcal{E} \to \mathcal{D}$, 
provided it exists, is the function such that $h^{-1}(h(x))=x$ for all $x\in \mathcal{D}$.
Given a function $h:\mathcal{D} \to \mathcal{E}$ taking on scalar (matrix) values, we denote 
by $h(x)^{-1}$ the reciprocal (inverse) of $h(x)$ at $x$. 
By \emph{coordinate transformation} $\tau:\mathcal{D} \to \mathcal{E}$ 
it is meant a local diffeomorphism \cite[p.~11]{isidori1995book}, that is
$\tau$ is a bijection and both $\tau$ and $\tau^{-1}$ are smooth mappings. A sufficient condition to check whether or not the function $\tau(x)$ defined on $\mathcal{D}$ is a local diffeomorphism is that the Jacobian matrix of $\tau$ is nonsingular at an interior point $x^0$ of $\mathcal{D}$. If this holds, then on an open subset $\mathcal{D}'$ of $\mathcal{D}$ containing $x^0$, $\tau(x)$ is a local diffeomorphism \cite[Proposition 1.2.3]{isidori1995book}.
We will also consider the notions of (vector) \emph{relative degree}
and \emph{Lie derivative}. The reader is referred to \cite[p.~220, 496]{isidori1995book} 
for their definitions and properties.

\section{Problem setting}
\label{Sec: PS}

\subsection{Feedback linearization problem}

Consider a continuous-time nonlinear system  
\begin{equation} \label{eq:system}
\dot{x}=f(x)+g(x)u
\end{equation}
where $x \in \mathbb{R}^n$ is the state, $u \in \mathbb{R}^m$ is
the input, $f: \mathbb{R}^n \to \mathbb{R}^n$, and
$g: \mathbb{R}^n \to \mathbb{R}^m$ are smooth vector fields.
The objective is to stabilize the system at some point $x=x^0$ using the so-called \emph{feedback linearization}
\cite[Chapter~5]{isidori1995book}, as we make precise in the sequel, in the case in which the vector fields 
$f(x), g(x)$ are not known but some priors about them are available.

We begin by recalling the \emph{state space exact linearization problem} or 
\emph{linearization problem}, for short \cite[p.~228]{isidori1995book}. 
Given a state $x^0$, find a neighborhood $\mathcal{D}$ of $x^0$, functions 
$\alpha\colon \mathcal{D}\to  \mathbb{R}^m$,  $\beta\colon \mathcal{D}\to  \mathbb{R}^{m\times m}$, 
a coordinate transformation
 $\tau\colon \mathcal{D}\to  \mathbb{R}^n$
and a controllable pair $(A,B)$ such that, for each $x \in \mathcal{D}$,
\begin{subequations} \label{eq:SSELP}
\begin{alignat}{2}
\displaystyle
\frac{\partial \tau}{\partial x} (f(x)+g(x)\alpha(x)) &= A \tau(x), \label{eq:SSELPa} \\[0.1cm]
\displaystyle
\frac{\partial \tau}{\partial x}g(x)\beta(x) &= B. \label{eq:SSELPb}
\end{alignat}
\end{subequations}

We slightly reformulate the problem in a form that is more convenient for our purposes.

\begin{lem} \label{lem:SSELP}
The linearization problem is solvable if there exist
a neighborhood $\mathcal{D}$ of $x^0$, functions $\gamma\colon \mathcal{D}\to  \mathbb{R}^{m\times m}$,   
$\delta\colon \mathcal{D}\to  \mathbb{R}^m$, 
with $\gamma(x)$ nonsingular for all $x\in \mathcal{D}$, 
and a coordinate transformation $\tau\colon \mathcal{D}\to  \mathbb{R}^n$ such that,
for each $x\in \mathcal{D}$ and $u\in \mathbb{R}^m$, \smallskip
\begin{equation}
\label{syst.after.change.coord}
\displaystyle\frac{\partial \tau}{\partial x} (f(x)+g(x)u)= A_c \tau(x) + B_c(\delta(x) + \gamma(x)u)
\end{equation}
\smallskip
where 
$A_c=\text{diag}(A_1, \ldots, A_m)$, $B_c=\text{diag}(B_1, \ldots, B_m)$,
\begin{equation}\label{brun-form}
A_i =\begin{bmatrix}
0 &1 & 0 & \ldots & 0\\
0 &0 & 1 & \ldots & 0\\
\vdots & \vdots & \vdots & \ddots &\vdots \\
0 &0 & 0 & \ldots & 1\\
0 &0 & 0 & \ldots & 0
\end{bmatrix}\in \mathbb{R}^{r_i\times r_i}, \quad
B_i =\begin{bmatrix}
0 \\
0\\
\vdots\\
0 \\
1
\end{bmatrix}\in \mathbb{R}^{r_i\times 1},
\end{equation}
with $r_1, \ldots, r_m$ nonnegative integers such that $r_1+\ldots +r_m=n$. 

Further, if $g(x^0)$ has rank $m$, then \eqref{syst.after.change.coord}
is also necessary. 
\qedp
\end{lem}
We refer the reader to \cite[section 5.2]{isidori1995book} for a proof.
By Lemma \ref{lem:SSELP}, condition \eqref{syst.after.change.coord} implies condition \eqref{eq:SSELP}
and is equivalent to \eqref{eq:SSELP} whenever $g(x^0)$ has full column rank. The latter requirement is
in fact quite mild: for single-input systems, it amounts to requiring that $g(x^0)$ is a nonzero vector.
It is convenient to focus on \eqref{syst.after.change.coord} 
as it expresses the linearization condition through one single equation that involves 
the flow dynamics $f(x)+g(x)u$. This turns out to be useful when we want to 
solve the linearization problem with data, as we can use data (trajectories)
as a proxy for $f(x)+g(x)u$. 

\subsection{Objective of the paper and assumptions}

Our objective is to discuss how to solve the linearization problem, i.e. how we can find
functions $\tau,\delta$ and $\gamma$ satisfying \eqref{syst.after.change.coord}. We will 
first discuss in Section \ref{sec:MB} the case where the dynamics of the system are known.
Then, in Section \ref{sec:DD}, we will discuss how we can approach the problem when
the dynamics of the system are unknown. 
Clearly, if the dynamics of the system are known, we can deal with the linearization problem using 
classical results \cite[Theorem 5.2.3]{isidori1995book}. The motivation to consider the case of 
known dynamics in  Section \ref{sec:MB} is that it introduces the line of thought that will be also used  
in Section \ref{sec:DD} when the dynamics are unknown.
The main assumptions in this work are:
\begin{assumption} \label{ass:elp-solvable}
The linearization problem is solvable. 
\qedp
\end{assumption}

\begin{assumption} \label{ass:library}
We know functions $Z: \mathcal{D} \to \mathbb R^{s}$,
$Y: \mathcal{D} \to \mathbb R^{p}$, and $W: \mathcal{D} \to \mathbb R^{r \times m}$, 
such that 
\begin{equation}
\tau(x)=\overline T Z(x),\, \delta(x)=\overline N Y(x),\, \gamma(x)=\overline M W(x)
\end{equation}
for some (unknown) 
matrices $\overline T$, $\overline N$, $\overline M$. 
\qedp
\end{assumption}
\begin{assumption} \label{ass:lin-independence}
The functions in $Z(x)$ ($Y(x)$, $W(x)$) are linearly independent functions on some neighborhood $\mathcal{D}'\subseteq \mathcal{D}$ of $x^0$. Namely, if 
$\sigma^\top Z(x)=0$ ($\rho^\top Y(x)=0$, $\pi^\top W(x)=0$) for all $x\in \mathcal{D}'$, then $\sigma=0$ ($\rho=0$, $\pi=0$). 
\qedp
\end{assumption}

An additional assumption will be introduced later.
Assumption \ref{ass:library} can be interpreted  by saying 
that we know a library of functions that includes the 
real system in the $\tau$-coordinates. We thus
require that some prior knowledge about the system is available, as it is the case 
with mechanical and electrical systems where some information about the
dynamics can be derived from first principles.
Assumptions of this type 
have been widely adopted in the context of system identification, see for 
example \cite{Brunton2016}, and recently also in the context of 
direct data-driven control, see \cite{dai2021statedependent,NonlinearityCancellation2023}.
We allow $Z,W,Y$ to contain terms not present in 
$\tau,\gamma,\alpha$, which accounts for an imprecise knowledge of 
the dynamics of interest (in fact, we may well have $s \gg n$ and $p,r \gg m$).
Thus, the goal is to discover the elements of the functions $Z,W,Y$ to obtain the identity \eqref{syst.after.change.coord}.  Assumption \ref{ass:lin-independence} is a natural assumption that excludes redundancy in the dictionaries chosen to represent the change of coordinates $\tau(x)$ and functions $\delta(x), \gamma(x)$. 

Before proceeding we make one final remark.
When \eqref{syst.after.change.coord} holds, 
a linearizing control law is $u=\gamma(x)^{-1}(v-\delta(x))$, with $v$ an arbitrary  signal. This yields
$\dot{\eta} = A_c \eta + B_cv$, where $\eta=\tau(x)$. 
Since $(A_c, B_c)$ is controllable, we can thus select 
$v = K\tau(x)$ with $K$ a state-feedback matrix that renders $A_c+B_cK$ Hurwitz.
As $\tau$ has a continuous inverse, with the extra property $\tau(x^0)=0$,
we will then have that $u=\gamma(x)^{-1}(K\tau(x)-\delta(x))$ stabilizes the nonlinear system at $x^0$.
While the design of $u$ is the obvious ultimate goal, it is important to stress
that \eqref{syst.after.change.coord} does \emph{not} involve the design of $K$;
it rather defines an intermediate step after which $K$, and thus $u$, can be readily designed. 
This point is crucial for the approach considered in the paper in the sense that 
by looking at \eqref{syst.after.change.coord} --without $K$-- we can obtain
a numerically efficient method, as detailed next.

\section{Main results}

We begin by expressing condition \eqref{syst.after.change.coord} in a convenient form.
Under Assumption \ref{ass:library}, condition \eqref{syst.after.change.coord} can be written as 
\begin{equation} \label{eq:MB_solution}
\overline T \displaystyle\frac{\partial Z}{\partial x} (f(x)+g(x)u)= 
A_c \overline T Z(x) + B_c(\overline N Y(x)+ \overline M W(x)u).
\end{equation}
Defining 
\begin{subequations} 
\label{eq:MB_solution_1}
\begin{alignat}{3}
& \ell_1(x,\dot{x}) := Z(x)^\top \otimes A_c - \left(\frac{\partial Z}{\partial x} \dot{x}\right)^\top \otimes I_n, \label{eq:MB_solution_1a}\\
& \ell_2(x) := Y(x)^\top \otimes B_c,\\
& \ell_3(x,u) := (W(x) u)^\top \otimes B_c,
\end{alignat}
\end{subequations}
and recalling the properties of the vectorization operator,
\eqref{eq:MB_solution} can be thus given the equivalent form 
\begin{equation} \label{identity}
F(x,u, \dot x)\overline v=0,
\end{equation}
where
\begin{subequations}
\begin{alignat}{3}
F(x,u, \dot x)
&= 
\begin{bmatrix}
\ell_1(x,\dot{x}) & \ell_2(x) & \ell_3(x,u) 
\end{bmatrix},\label{F} \\[2mm] 
\overline v &= 
\begin{bmatrix}
{\rm vec}(\overline T)\\
{\rm vec}(\overline N)\\
{\rm vec}(\overline M)\
\end{bmatrix}. \label{eq_overline_v}
\end{alignat}
\end{subequations}
By Assumptions  \ref{ass:elp-solvable} and \ref{ass:library}, there exists $\overline v\ne 0$ 
such that  \eqref{identity} holds for all $x\in \mathcal{D}$ and all $u\in \mathbb{R}^m$. 
The property $\overline v\ne 0$ holds because otherwise $\tau(x)$ would not be a change 
of coordinates and $\gamma(x)$ would not be nonsingular. 
Having assumed that the linearization problem is feasible and
bearing in mind that its solution takes the form \eqref{identity},
we thus focus on the problem of finding one solution of such a form, i.e., 
$v \ne 0$
such that 
\begin{equation}
\label{form-feedb-linerz-pbm1}
\begin{array}{l}
\hspace{-1.5mm}F(x,u, \dot x)v=0, \\[1mm]
\hspace{-1.5mm}\textrm{for all $x\in \mathcal{D}$, $u\in \mathbb{R}^m$, and $\dot x$ such that $\dot x=f(x)+g(x)u$}.
\end{array}
\end{equation}
The solution to this problem will be investigated in the next section. Once nonzero vectors $v$ that satisfy \eqref{form-feedb-linerz-pbm1} are found, we will need to make sure that these solutions include those for which the matrices  $(T,M,N)={\rm vec}^{-1}(v)$ additionally satisfy that $TZ(x)$ is a change of coordinates and $MW(x)$ is nonsingular in a neighborhood of $x^0$. This will be discussed in Section \ref{sec:space-F(D)=0}. 

\subsection{Model-based solution} \label{sec:MB}

We first derive a solution to the linearization problem assuming that the dynamics of the system are known.
This will serve as a basis for the case where the dynamics is unknown and we only have access 
to input-state data collected from the system.

We approach this problem in the following way. 
Since $f,g$ are known, the matrix of functions  
$F(x,u, f(x)+g(x)u))$ is also known. Let $F_{ij}$ denote the $(i,j)$-th entry of the matrix $F$. Knowing the functions $F_{ij}(x,u, f(x)+g(x)u))$ we can determine a set of linearly independent functions
over $\mathbb{R}$ defined on $\mathcal{D}\times \mathbb{R}^m$, 
$\{\phi_{k}(x,u)\}_{k=1}^{n_b}$, such that each $F_{ij}(x,u, f(x)+g(x)u))$ 
can be expressed as a linear combination of this basis, that is
\be\label{lin.comb}
F_{ij}(x,u, f(x)+g(x)u))= \phi(x,u)^\top  c_{ij},
\ee 
\be\label{vector.basis.functions}
\phi(x,u):=\begin{bmatrix}
\phi_{1}(x,u)& \ldots & \phi_{n_b}(x,u)
\end{bmatrix}^\top,
\ee 
and $c_{ij}\in \mathbb{R}^{n_b}$ is a vector of coefficients.\footnote{Similar to Assumption \ref{ass:lin-independence}, the notion of linearly independent functions that we adopt here is as follows: there exists a vector of coefficients $c\in \mathbb{R}^{n_b}$ such  that $c^\top \phi(x,u)$ is the identically zero function if and only if $c=0$.} Notice that 
the subscript of $c_{ij}$ is used to match the indexing of $F_{ij}$, and not to indicate some entry of the vector $c$.
Here, each $c_{ij}$ is unique because $\phi(x,u)$ consists of independent functions.
\begin{theorem}\label{coord.transf.model.based}{\bf \emph{(Model-based solution with complete dictionaries)}}
Let Assumptions \ref{ass:elp-solvable} and \ref{ass:library} hold.
Let $F(x,u,\dot{x})$ be as in \eqref{F}.
Let $\{\phi_{k}(x,u)\}_{k=1}^{n_b}$ be a set of linearly independent functions over 
$\mathbb{R}$  defined on the set $\mathcal{D}\times \mathbb{R}^m$, such that each entry $F_{ij}$ of $F$ is expressed as in \eqref{lin.comb}-\eqref{vector.basis.functions} on $\mathcal{D}\times \mathbb{R}^m$. 
Then,  
$v$
satisfies \eqref{form-feedb-linerz-pbm1} if and only if 
$v$ is a  solution of 
the system of 
$n n_b$ linear  equations
\begin{equation} \label{eq:sol:MB}
\begin{bmatrix}
c_{11} & c_{12}  & \ldots & c_{1\mu}\\[0.1cm]
\vdots & \vdots & \ddots & \vdots\\[0.1cm]
c_{n1} & c_{n2} & \ldots & c_{n\mu}\\
\end{bmatrix}
v=0
\end{equation}
where 
$\mu = ns + pm + rm$ denotes the size of $v$. 
 \qeds
\end{theorem}

\emph{Proof.} Write the system of equations 
$F(x,u, f(x)+g(x)u))v=0$ in \eqref{form-feedb-linerz-pbm1} row by row as
\begin{equation}
\sum_{j=1}^{\mu} F_{ij}(x,u, f(x)+g(x)u)v_j=0,
\quad i=1,2,\ldots,n
\end{equation}
By \eqref{lin.comb}, the equations above can be re-written as
\begin{equation}
\begin{array}{ll}
& \displaystyle \sum_{j=1}^{\mu} F_{ij}(x,u, f(x)+g(x)u)v_j  \\[0.2cm]
& = \displaystyle \sum_{j=1}^{\mu} \phi(x,u)^\top  c_{ij} v_j  \\[0.2cm]
& = \phi(x,u)^\top \displaystyle \sum_{j=1}^{\mu}   c_{ij} e_j^\top v=0, \quad i=1,2,\ldots,n,
\end{array}
\end{equation}
where $e_j$ denotes the $j$-th vector of the canonical basis of $\mathbb{R}^{\mu}$.
As the vector $\phi(x,u)^\top$ is made of linearly independent functions over $\mathbb{R}$, the equation  
\begin{equation}
\phi(x,u)^\top \sum_{j=1}^{\mu}   c_{ij} e_j^\top v=0
\end{equation}
holds if and only if $\sum_{j=1}^{\mu}   c_{ij} e_j^\top v=0$. 
This equation must hold for every integer $i$. We conclude that  the set of all  the 
vectors 
$v$ such that \eqref{form-feedb-linerz-pbm1} holds is given by the solutions to the system of linear equations \eqref{eq:sol:MB}.  \qedp

A solution of interest is obtained by excluding the solution $v=0$ from the set 
of solutions to the system of linear equations. One can further seek among all these 
solutions at least one, say $\hat v$, for which 
$\hat T \frac{\partial Z}{\partial x} (x^0)$ and $\hat M W(x^0)$ are nonsingular. 
\begin{example} {\bf \emph{(Model-based feedback linearization)
}}\label{ex:MB}
Consider the nonlinear system \cite{Kaiser2021}
\begin{equation} \label{eq:example_1}
\left\{ \begin{array}{l}
\dot{x}_1 = \mu x_1 + u \\[0.1cm]
\dot{x}_2 = \lambda (x_2 - x_1^2) + u,
\end{array}
\right.
\end{equation} 
where $\mu \neq \lambda$ and
where the equilibrium of interest is $x^0=0$. 
The system satisfies the necessary and sufficient conditions for the linearization problem 
to be solvable \cite[Theorem 4.2.3]{isidori1995book}. 
To determine the coordinate transformation, one could solve the partial differential equation 
$\frac{\partial \lambda}{\partial x} g(x) =0 $
with the nontriviality condition $\frac{\partial \lambda}{\partial x}[f,g](0)\ne 0$, where 
$[\cdot,\cdot]$ denotes the Lie bracket. Once $\lambda$ is obtained, 
the coordinate transformation $\tau$ would be obtained setting 
$\tau(x)=[\begin{smallmatrix} \lambda(x) & \frac{\partial \lambda}{\partial x} f(x) \end{smallmatrix}]^\top$. 
Here, we pursue the approach outlined in Theorem \ref{coord.transf.model.based}. 
Suppose that we choose the following library of basis functions: 
\begin{equation} \label{eq:example_basis}
Z(x) = \begin{bmatrix} x \\ x_1^2 \\ x_2^2 \end{bmatrix}, \quad
Y(x) = Z(x), \quad
W(x) = \begin{bmatrix} 1 \\ Z(x) \end{bmatrix}.
\end{equation}
We add the constant factor $1$ in $W(x)$ to  make sure that 
$\gamma(x)= \overline{M} W(x)$ is nonsingular around $x^0=0$. 
With this choice, $F(x,u, \dot x)$ is made of the following sub-matrices:
{\setlength\arraycolsep{1pt} 
\begin{subequations}
\begin{alignat}{5}
& \ell_1(x,\dot x) = \nonumber \\
& \,\,\,\, \begin{bmatrix}
-\dot x_1 & x_1 & -\dot x_2 & x_2 & -2x_1 \dot x_1 & x_1^2 & -2x_2 \dot x_2 & x_2^2
\\ 0 & -\dot x_1 & 0 & -\dot x_2 & 0 & -2x_1 \dot x_1 & 0 & -2x_2 \dot x_2
\end{bmatrix}, \\
& \ell_2(x)=\setlength\arraycolsep{2.5pt}
\begin{bmatrix}
0 & 0 &0 &0\\
x_1 & x_2 & x_1^2 & x_2^2
\end{bmatrix}, \, \\ &\ell_3(x,u)=\setlength\arraycolsep{2.5pt} 
\begin{bmatrix}
0 &0 & 0 &0 &0\\
u & x_1 u & x_2 u & x_1^2 u  & x_2^2 u 
\end{bmatrix}.
\end{alignat}
\end{subequations}}%
Note that ${\mu={\rm size}(v)=17}$. Bearing in mind the expression of $\dot x_1, \dot x_2$, 
the basis functions $\{\phi_{k}(x,u)\}_{k=1}^{n_b}$ in terms of which the entries of $F(x,u, \dot x)$ can be expressed are 
\begin{equation}
x_1,\, x_2,\,  u,\,  x_1^2,\,  x_2^2,\,  x_1u,\,  x_2 u,\,  x_1^2 x_2,\,  x_1^2 u,\,  x_2^2 u .
\end{equation}
Hence, $n_b=10$.  This choice gives a unique set of coefficients $c_{ij}$, with $i=1,2$ and 
$j=1,2,\ldots, {\mu}$. which, once replaced in \eqref{eq:sol:MB}, 
gives rise to the following system of equations:
\begin{equation}
\begin{array}{rr}
- \mu v_1 +v_2= 0,& -\lambda v_3 +v_4=0,\\
\lambda v_3 -2\mu v_5 + v_6 =0, & -2\lambda v_7 +v_8=0, \\
-\mu v_2 +v_9= 0,& -\lambda v_4 +v_{10}=0,\\
-v_2 -v_4+v_{13}=0, & \! \! \! \!\lambda v_4-2\mu v_6 +v_{11}=0,\\
-2v_6+v_{14}=0, & 2v_8=v_{15}=0, \\
v_1 +v_3=0, & v_5=v_7=0, \\
 -2\lambda v_8+v_{12}=0, & v_{16}=v_{17}=0,
\end{array}
\end{equation}
whose solutions are given by 
{\setlength\arraycolsep{3.4pt}
\begin{equation}
\hspace{-3mm}\begin{array}{r}
v^\top = \big[ 
\begin{matrix}
v_1 & \mu v_1 & -v_1 &  -\lambda v_1 &  0 &  \lambda v_1 &  0 &  0 & \mu^2 v_1 & -\lambda^2 v_1
\end{matrix} \\
\begin{matrix}
\lambda (2\mu+\lambda)v_1 & 0 &  (\mu-\lambda)v_1 & 2\lambda v_1 & 0 & 0 & 0
\end{matrix}
\big],
\end{array}
\end{equation}}%
with $v_1$ a free parameter. The first $8$ entries of $v$ define $\overline T$:
\begin{equation}
\overline T= v_1 
\begin{bmatrix}
1 & -1 & 0 & 0\\
\mu & -\lambda & \lambda & 0
\end{bmatrix}
\end{equation}
which returns the change of coordinates 
\begin{equation}
\tau: x \mapsto 
v_1
\begin{bmatrix}
x_1 - x_2\\
\mu x_1 -\lambda (x_2 - x_1^2)
\end{bmatrix}
\end{equation}
The $9$th--$12$th entries of $v$ define $\overline N$ and return 
$\delta(x)=\overline N Y(x)=v_1(\mu^2 x_1  -\lambda^2 x_2 +\lambda(2\mu+\lambda)x_1^2)$.
The $13$th--$17$th entries  define $\overline M$ and return 
$\gamma(x)=\overline M W(x)=(\mu-\lambda)v_1 +2\lambda v_1 x_1$. For any 
$v_1\ne 0$, we obtain feasible $\tau(x), \gamma(x), \delta(x)$. 
The solution coincides with the one guaranteed by \cite[Theorem 4.2.3]{isidori1995book}. 
\qedp
\end{example}

\subsection{Data-based solution and generalization from the 
sample space} \label{sec:DD}

The approach described in the previous section requires 
the knowledge of the vector fields $f,g$. We can draw inspiration from 
the model-based approach to find a solution achievable using data alone.
We start again from the equation $F(x,u, \dot x)v=0$, but instead of expressing $F(x,u, \dot x)$ 
via the basis functions $\{\phi_{k}(x,u)\}_{k=1}^{n_b}$, which is not possible since the vector fields 
defining  $ \dot x$ are unknown, we evaluate $F(x,u, \dot x)$ on a dataset and use this information 
to build a solution $v$ to \eqref{form-feedb-linerz-pbm1} when $f,g$ are unknown. 

Namely, assume that is is possible to make an experiment on the 
system and collect a dataset 
\be\label{dataset}
\mathbb{D}:=\{(x_i, u_i, \dot{x}_i)\}_{i=0}^{L-1},
\ee
with $x_i \in \mathcal{D}$, $u\in \mathbb{R}^m$, and $\dot x_i= f(x_i)+g(x_i)u_i$.
The idea is then to solve \eqref{form-feedb-linerz-pbm1} at the collected data points,
namely to find a vector $v\ne 0$ belonging to the kernel of $\mathcal{F}(\mathbb{D})$ where
\begin{equation} \label{eq:mathcalF}
\mathcal{F}(\mathbb{D}):= 
\begin{bmatrix}
F(x_0,u_0,\dot x_0)\\
\vdots
\\
F(x_{L-1},u_{L-1},\dot x_{L-1})\\
\end{bmatrix}.
\end{equation}
Note
that each entry of $\mathcal{F}(\mathbb{D})$ is of the form $F(x,u,\dot{x})$
with $\dot{x_i}=f(x_i)+g(x_i)u_i$ for each $i=0,\ldots,L-1$.

The matrix $\mathcal{F}(\mathbb{D})$ can be computed from data, after which 
we easily determine its null space. This procedure, however, 
only guarantees that the dynamics are linearized
at the collected data points, i.e., in the \emph{sample} space. Clearly, we would like that 
to ascertain whether a solution defines a coordinate transformation that is valid on a whole space
of interest, and this involves the problem of generalizing from a \emph{finite} set of data points 
to \emph{infinitely} many data points. It is actually possible to give a simple condition that resolves
this issue. 
The condition amounts to require the dataset to be sufficiently rich in the sense specified as follows. 
\begin{assumption}\label{asspt:data-rich}
Consider the basis  functions in \eqref{vector.basis.functions}. The samples $\{(x_i, u_i)\}_{i=0}^{L-1}$ obtained from the dataset $\mathbb{D}$ in \eqref{dataset} are such that 
\be\label{data-rich-cond}
{\rm rank} 
\begin{bmatrix}
\phi(x_0,u_0)^\top\\
\vdots\\
\phi(x_{L-1},u_{L-1})^\top\\
\end{bmatrix}
=
n_b. \qquad \qquad\text{{\tiny $\blacksquare$}}
\ee
\end{assumption}

\medskip

Note that this assumption requires the number of samples $L$ to be not less than the number $n_b$ of functions in the basis. 

\medskip

\begin{theorem}\label{thm:DD-2}{\bf \emph{(Data-driven feedback linearization)}}
Let Assumptions \ref{ass:elp-solvable} and \ref{ass:library} hold. The following hold:
\begin{enumerate}[(i)]
\item 
If 
$v$
satisfies \eqref{form-feedb-linerz-pbm1}, then $v\in \ker (\mathcal{F}(\mathbb{D}))$. 
\item Let Assumption  \ref{asspt:data-rich} hold.  If 
$v\in \ker (\mathcal{F}(\mathbb{D}))$, 
then $v$ 
satisfies \eqref{form-feedb-linerz-pbm1}.  
\qeds
\end{enumerate}
\end{theorem}

{\it Proof.} (i)  holds because, by definition,  $v$ satisfies \eqref{form-feedb-linerz-pbm1} if and only if $F(x, \dot x, u)v=0$  for all $x\in \mathcal{D}$, all $u\in \mathbb{R}^m$, where $\dot x = f(x)+g(x)u$. The equality holds in particular at the samples in $\mathbb{D}$, hence  $F(x_i, \dot x_i, u_i)v=0$ for $i=1,2,\ldots, L$. By definition of $\mathcal{F}(\mathbb{D})$, this implies  that $v\in \ker(\mathcal{F}(\mathbb{D}))$. 

(ii) From the proof of Theorem \ref{coord.transf.model.based}, it is  known that 
\be\label{form.proof.thm1}
\ba{rl}
&
F(x, f(x)+g(x)u, u)  \\
=& \begin{bmatrix}
\phi(x,u)^\top c_{11}& \ldots & \phi(x,u)^\top c_{1\mu}\\
\vdots & \ddots & \vdots\\
\phi(x,u)^\top c_{n1}& \ldots & \phi(x,u)^\top c_{n\mu}
\end{bmatrix}
\\
=& 
\underbrace{
\begin{bmatrix}
\phi(x,u)^\top  & \ldots &  \mathbb{0} \\
\vdots & \ddots & \vdots\\
\mathbb{0}& \ldots & \phi(x,u)^\top
\end{bmatrix}
}_{I_n\otimes \phi(x,u)^\top}
\underbrace{
\begin{bmatrix}
\sum_{j=1}^{\mu}   c_{1j} e_j^\top\\[0.1cm]
\vdots\\[0.1cm]
\sum_{j=1}^{\mu}   c_{nj} e_j^\top\\
\end{bmatrix}
}_{=:\Gamma} 
\ea 
\ee
where, for all $i=1,2,\ldots, n$ and all $j=1,2,\ldots, \mu$,  the vector $c_{ij}\in \mathbb{R}^{n_b}$ is such that $F_{ij}(x, f(x)+g(x)u, u)=\phi(x,u)^\top c_{ij}$ and $e_j\in \mathbb{R}^{\mu}$ is the $j$th vector of the canonical base. Set 
$\Phi(x, u):= I_n\otimes \phi(x, u)^\top$ for the sake of brevity. 
Then 
\be\label{F.Phi.Gamma}
F(x, f(x)+g(x)u, u)= \Phi(x,u)\Gamma
\ee
By the definition \eqref{eq:mathcalF} of $\mathcal{F}(\mathbb{D})$, the latter can be written as 
\[
\mathcal{F}(\mathbb{D})= 
\begin{bmatrix}
\Phi(x_0, u_0)\\
\vdots\\
\Phi(x_{L-1}, u_{L-1})
\end{bmatrix}\Gamma.
\]
Observe that
\[
\begin{bmatrix}
\phi(x_0,u_0)^\top\\
\vdots\\
\phi(x_{L-1},u_{L-1})^\top\\
\end{bmatrix}
\]
has full column rank, i.e., condition \eqref{data-rich-cond} holds, if and only if\footnote{See Lemma \ref{lem:renk-equivalence} in the Appendix for details.}  the matrix
\[
\begin{bmatrix}
\Phi(x_0, u_0)\\
\ldots\\
\Phi(x_{L-1}, u_{L-1})
\end{bmatrix}
\]
has full column rank. Hence, under condition \eqref{data-rich-cond}, $v \in \ker(\mathcal{F}(\mathbb{D}))$ if and only if $v\in \ker (\Gamma)$. But then, in view of \eqref{F.Phi.Gamma},   under condition \eqref{data-rich-cond}, $v \in \ker(\mathcal{F}(\mathbb{D}))$ implies $F(x, u, f(x)+g(x)u)v=0$. This ends the proof. \qedp
\medskip

Condition \eqref{data-rich-cond} is a mild assumption on the dataset, which in some cases  can  be abandoned, as pointed out in the following result: 
\begin{proposition}\label{prop:nullity=1} Let ${\rm nullity}(\mathcal{F}(\mathbb{D}))=1$, where ${\rm nullity}(M)$ denotes the dimension of the null space of $M$. Then, any
vector $v\ne 0$ belonging to $\ker(\mathcal{F}(\mathbb{D}))$  satisfies
(10). 
\qeds
\end{proposition}

{\em Proof.} See \cite[Theorem 2]{de2023complete} for a proof. \qedp

\medskip

\begin{example}{\bf \emph{(Data-driven feedback linearization with ${\rm nullity}(\mathcal{F}(\mathbb{D}))=1$) }}
\label{ex:DD}
Consider again the nonlinear system 
\begin{equation} \label{eq:example_1a}
\left\{ \begin{array}{l}
\dot{x}_1 = \mu x_1 + u \\[0.1cm]
\dot{x}_2 = \lambda (x_2 - x_1^2) + u,
\end{array}
\right.
\end{equation} 
where the equilibrium of interest is $x^0=0$.
We assume $\mu=-0.5$ and $\lambda=0.2$.
According to Example \ref{ex:MB},
\begin{equation} \label{eq:example_coord_change}
\tau: x \mapsto \begin{bmatrix} x_1-x_2 \\ -0.5x_1-0.2(x_2-x_1^2) \end{bmatrix}
\end{equation} 
defines a change of variables linearizing the system about the origin. 
Specifically, in the coordinates $\eta=\tau(x)$ we have
\begin{equation} \label{eq:example_new_coord}
\left\{ \begin{array}{l}
\dot{\eta}_1 = \eta_2 \\[0.1cm]
\dot{\eta}_2 = \underbrace{0.25x_1-0.04x_2-0.16x_1^2}_{\delta(x)} + \underbrace{(0.4 x_1-0.7)}_{\gamma(x)} u.
\end{array}
\right.\end{equation} 
Suppose that the model is unknown and we want to discover this coordinate transformation using 
data collected from the system. We make an experiment on the 
system of duration $10$s in which we apply a piecewise constant
input uniformly distributed within $[-0.1,0.1]$ and with initial conditions
in the same interval. We collect $L=100$ samples $\{x_i,u_i,\dot{x}_i\}$ 
with period $0.1s$. Suppose that 
we choose the following library of basis functions: 
\begin{equation} \label{eq:example_basis_E2}
Z(x) = \begin{bmatrix} x \\ x^2 \\ x^3 \\ \sin(x) \\ \cos(x) \end{bmatrix}, \quad
Y(x) = Z(x) , \quad 
W(x) = \begin{bmatrix} 1 \\ Z(x) \end{bmatrix},
\end{equation} 
where by $x^2$ we mean the vector with components $x_1^2$ and $x_2^2$, and 
the same meaning holds for the sine, the cosine and the cubic function. 
Here the library is richer than the one of Example \ref{ex:MB} to illustrate the
situation where we add several candidate functions to
compensate the lack of knowledge of the dynamics of the system. 
Using this library and the dataset, we compute $\mathcal{F}(\mathbb{D})$ and determine its null space.
The dimension of the null space of $\mathcal{F}(\mathbb{D})$ is $1$, and the solution is indeed as in 
\eqref{eq:example_solution} (modulo a constant factor)
\begin{subequations} \label{eq:example_solution}
\begin{alignat}{99}
&\overline{T} = \begin{bmatrix} 1 & -1 & 0 & 0 & 0 & 0 & 0 & 0 & 0 & 0 \\ 
-0.5 & -0.2 & 0.2 & 0 & 0 & 0 & 0 & 0 & 0 & 0 \end{bmatrix} \\
&\overline{N} = \begin{bmatrix} 0.25 & -0.04 & -0.16 & 0 & 0 & 0 & 0 & 0  & 0 & 0 \end{bmatrix} \\
& \overline{M} = \left[ \begin{array}{ccccccccccc}
-0.7 &  0.4 & 0 & 0 & 0 & 0 & 0 & 0 & 0 & 0 & 0 \end{array} \right],
\end{alignat}
\end{subequations} 
as certified by Proposition \ref{prop:nullity=1}.
One sees that the algorithm automatically discards sine, cosine and cubic functions,
recovering the model-based solution.
In this example, we systematically obtained the same solution with different datasets
generated considering different input signals. 
\qedp
\end{example}

We conclude this section with some remarks discussing theoretical and  
practical aspects of the data-based solution.

\begin{remark}{\bf \emph{(Domain of linearization and region of attraction (RoA))}}
  Assumption \ref{ass:library} requires that we know the domain of
  linearization $\mathcal{D}$.  In practice, this can be inferred from
  data by determining the domain over which $TZ(x)$ is a
  diffeomorphism. It is interesting to note that the knowledge of
  $\mathcal{D}$ permits us to immediately obtain an estimate of the
  RoA. In fact, once we have the linear dynamics
  $\dot{\eta}=A_c \eta + B_c v$ in the coordinates $\eta=\tau(x)$ we
  can determine a stabilizing control law $v=K\tau(x)$ and a RoA, say
  $\mathcal{S}$, for the system $\dot{\eta}=(A_c+B_cK)\eta$ that is
  also an invariant set (for instance, a Lyapunov sublevel set).
  Thus, a RoA for the nonlinear system can be obtained as any
  invariant set $\mathcal{R} \subseteq \mathcal{S}$ such that
  $\tau^{-1}(\eta) \in \mathcal{D}$ for all $\eta \in \mathcal{R}$.
  \qedp
\end{remark}

\begin{remark}{\bf \emph{(Choice of the library and noisy data)}}
  Our approach crucially depends on Assumption \ref{ass:library}. As
  shown in Example \ref{ex:DD}, we can be generous with the number of
  candidate functions as the solution is eventually obtained by
  determining the null space of a matrix ($\mathcal{F}(\mathbb{D})$),
  which is computationally fast even for big matrices. The situation
  can be different with noisy data.  The main issue with noisy data is
  that all the functions might be selected (the 
  overfitting
  problem).  In this case, some prior knowledge can help to keep the
  size of the library moderate.  In parallel, it can be useful to
  explicitly search for \emph{sparse} solutions, for example:
  $\text{minimize}_{v: \|v\|=1} \|\mathcal{F}(\mathbb{D}) v\|^2_2 +
  \alpha \| v \|_0$, where $\|\mathcal{F}(\mathbb{D}) v\|^2_2$
  replaces the constraint $\mathcal{F}(\mathbb{D}) v=0$, while
  $\alpha \|v\|_0$, $\alpha>0$, penalizes the complexity of the
  solution; finally, the constraint $\|v\|=1$ ensures that the
  solution is different from zero.  Formulations of this type have
  been widely adopted in the context of data-driven control \emph{cf.}
  \cite{NonlinearityCancellation2023,alsalti2021data,Brunton2016,bridging}.
  Treatment of this problem is left for future
  research. 
\qedp
\end{remark}

\section{Discussion}\label{ref:discussion}

\subsection{Basis functions}\label{subsec:basis-functs}
To check condition \eqref{data-rich-cond}, the basis functions $\{\phi_{k}(x,u)\}_{k=1}^{n_b}$ must be known. This information can be inferred from the dictionaries $Z(x)$, $W(x)$ provided that they have been chosen to satisfy Assumption \ref{ass:library} {\em and} to include the functions comprising the  vector fields $f(x)$ and $g(x)$ such that the latter can be expressed as $f(x)= A_* Z(x)$ and $g(x)= B_* W(x)$, where $A_*$ and $B_*$ are unknown matrices. In this case, the function $\ell_1$ in \eqref{eq:MB_solution_1a} becomes
\[\ba{ll}
\ell_1(x,u) := &Z(x)^\top \otimes A_c - \Bigl(({\rm vec}A_*)^\top \left(Z(x)\otimes \frac{\partial Z}{\partial x}^\top\right) \\
&
+ ({\rm vec}B_*)^\top \left(W(x)u\otimes \frac{\partial Z}{\partial x}^\top \Bigr)\otimes I_n\right),
\ea\]
where we stress  the dependence on $x,u$ and not $x, \dot x$. In fact, the entries of $\ell_1(x,u)$ are now linear combinations of the functions in $Z(x)$, $Z(x)\otimes \frac{\partial Z}{\partial x}^\top$, $W(x)u\otimes \frac{\partial Z}{\partial x}^\top$, which depend on $x,u$ and are {\em known}. The knowledge of all the functions in $\ell_1(x,u)$ (and  $\ell_2(x)$, $\ell_3(x,u)$) allows us to select the set of linear independent functions that form the basis $\{\phi_{k}(x,u)\}_{k=1}^{n_b}$. Knowing this basis, we can check whether or not \eqref{data-rich-cond} holds. 

A related problem is the one of designing an experiment that generates the dataset $\mathbb{D}$ such that the condition \eqref{data-rich-cond} holds. This is yet another instance of the problem of generating persistently exciting inputs that induce a sufficiently rich state response. Namely, one would like to design the input sequence $\{u_i\}_{i=0}^{L-1}$ such that the resulting $L$-long input-state dependent vector-valued signal $\{\hat \phi(x_i, u_i)\}_{i=0}^{L-1}$, where 
$\hat \phi(x_i, u_i):=\left[\begin{smallmatrix}  \phi_1(x_i, u_i)& \ldots & \phi_{n_b}(x_i, u_i) \end{smallmatrix}\right]^\top$, makes the Hankel matrix of depth $1$ $\left[\begin{smallmatrix}  \hat \phi(x_0, u_0) & \ldots & \hat \phi(x_{L-1}, u_{L-1})\end{smallmatrix}\right]$ a full-row rank matrix. For nonlinear systems such a problem has not yet been thoroughly investigated, although some results are available \cite{dpt2021design}, \cite{alsalti2023}.

\subsection{The space of solutions of $\mathcal{F}(\mathbb{D})v=0$}\label{sec:space-F(D)=0}
The set of solutions of $\mathcal{F}(\mathbb{D})v=0$ constitute a vector space of dimension $\nu\ge 1$ that depends on the choice of the dictionary. In fact, as we underscore in the next results, each vector  of the space of solutions of $\mathcal{F}(\mathbb{D})v=0$ necessarily determine  an 
``output" function  that (i)  is obtainable from linear combinations of the functions in $Z(x)$; (ii) satisfy certain invariance conditions that involve the functions in $Z(x)$, $Y(x)$, $W(x)$ (see \eqref{cond-rel-degree}, \eqref{inv-cond-onT1} below).
\footnote{ We state that the function $T_1^i Z(x)$, rather than the vector $T_1^i$, satisfies \eqref{cond-rel-degree}, \eqref{inv-cond-onT1} in view of the identities $T^i_{1} L_{g} L_f^{k}Z(x)=
L_{g} L_f^{k} T^i_{1} Z(x)$, $T^i_{1}L_f^{k}Z(x)=L_f^{k} T^i_{1} Z(x)$, for any $1\le i\le m$ and any $0\le k\le r_i -1$.} 
Conversely, any nontrivial $m$-tuple of 
``output" functions  obtainable from a linear combination of the functions in $Z(x)$ and satisfying \eqref{cond-rel-degree}, \eqref{inv-cond-onT1}, including those that solve the feedback linearization problem, determines a nonzero solution $v$ to  $\mathcal{F}(\mathbb{D})v=0$.



\begin{proposition}\label{lem:dimension-nu-space-V}
Let Assumptions \ref{ass:elp-solvable}--\ref{ass:lin-independence} hold. 
Consider any nonzero vector $v$ that satisfies $\mathcal{F}(\mathbb{D})v=0$ and let the matrices $(T,N,M)={\rm vec}^{-1}(v)$ be partitioned as 
\be\label{T-M-N}
T=\begin{bmatrix}
T^1_1\\
\vdots\\
T^1_{r_1}\\
\vdots\\
T^m_1\\
\vdots\\
T^m_{r_m}\\
\end{bmatrix}, 
M=\begin{bmatrix}
M_1\\
\vdots\\
M_m
\end{bmatrix}, 
N=\begin{bmatrix}
N_1\\
\vdots\\
N_m
\end{bmatrix},
\ee
where $m$ is the dimension of the input $u$ and $r_1, \ldots, r_m$ are the integers defined in \eqref{brun-form}. 
Then, there exists $i\in \{1, 2, \ldots, m\}$ such that $T_1^i \ne 0$ and the function $T_1^i Z(x)$
satisfies the conditions\footnote{By $L_{g} L_f^{j}Z(x)$ we are denoting the vector 
$[L_{g_1} L_f^{j}Z(x) \; \ldots \;L_{g_m} L_f^{j}Z(x)]$.
}
\be\label{cond-rel-degree}\ba{l}
T^i_{1} L_{g} Z(x)=T^i_{1} L_{g} L_f Z(x)=\ldots= T^i_{1} L_{g} L_f^{r_i-2}Z(x)=0,
\ea\ee
and
\be\label{inv-cond-onT1}
\!\!\!\!
\ba{rll}
T^i_{1} L_f Z(x), \ldots, T^i_{1}  L_f^{r_i-1}Z(x)&\!\!\!\!\!\!\in& \!\!\!\!\!\!{\rm span}_{\mathbb{R}}\{Z_1(x), \ldots, Z_s(x)\},\\[2mm]
T^i_{1}  L_f^{r_i}Z(x) &\!\!\!\!\!\!\in&\!\!\!\!\!\! {\rm span}_{\mathbb{R}}\{Y_1(x), \ldots, Y_p(x)\},\\[2mm]
T^i_{1}  L_{g_1} L_f^{r_i-1}Z(x)&\!\!\!\!\!\!\in&\!\!\!\!\!\!\!{\rm span}_{\mathbb{R}}\{W_{11}(x), \ldots, W_{r1}(x)\},\\
\vdots &\\
T^i_{1}  L_{g_m} L_f^{r_i-1}Z(x)&\!\!\!\!\!\!\in&\!\!\!\!\!\!{\rm span}_{\mathbb{R}}\{W_{1m}(x), \ldots, W_{rm}(x)\},
\ea
\ee
for all $x\in \mathcal{D}$, 
where the functions $Z_j(x)$, $1\le j\le s$, $Y_j(x)$, $1\le j\le p$,  $W_{ik}(x)$,  $1\le i\le r$,  $1\le k\le m$, are the entries of $Z(x)$, $Y(x)$, $W(x)$.  Moreover, for any $i\in \{1, 2, \ldots, m\}$ such that $T_1^i \ne 0$, the vectors $T^i_2,\ldots, T^i_{r_i}, N_i,M_i$ linearly depend on $T^i_1$. \qeds
\end{proposition}

{\it Proof.} The proof is deferred to Section \ref{sec:proof-prop-2}. 
%
\qedp 

The following converse result  holds. 

\begin{proposition}\label{prop3}
Let Assumptions \ref{ass:elp-solvable} and \ref{ass:library} hold. 
For every $i\in \{1,\ldots,m\}$ there exists $T^i_1\ne 0$ such that  \eqref{cond-rel-degree}, \eqref{inv-cond-onT1} hold. Moreover, there exist row vectors $T^i_2,\ldots, T^i_{r_i}, N_i,M_i$ depending on $T^i_1$ such that  ${\rm vec}(T,N,M)=:v$ satisfies 
$\mathcal{F}(\mathbb{D})v=0$. \qeds
\end{proposition}

{\em Proof.} See Section \ref{sec:proof-prop3}.  \qedp

Hence, all the $m$-tuple of output functions obtainable from $Z(x)$ via the coefficient vectors $T^1_1, \ldots, T^m_1$ and satisfying \eqref{cond-rel-degree}, \eqref{inv-cond-onT1} are included in the set of solutions of  $\mathcal{F}(\mathbb{D})v=0$. 
%
One also deduces that the dimension $\nu$ of the space of solutions $v$ to $\ker(\mathcal{F}(\mathbb{D}))$ cannot exceed the number of  entries of the vectors $T_1^1, \ldots, T_1^m$, which is $ms$, where $m$ is the dimension of the input space and $s$ is the number of functions of $Z(x)$. In fact, in view of the conditions \eqref{cond-rel-degree}, \eqref{inv-cond-onT1}, such a dimension will be in general smaller than $ms$. Moreover, as these conditions depend on $Z(x), W(x), Y(x)$, by changing the dictionaries, we might obtain a different space of solutions of $\mathcal{F}(\mathbb{D})v=0$. We illustrate these points with an  example.

%

\medskip

\begin{example}\label{empl:extended-Z}
{\bf \emph{(Data-driven feedback linearization with ${\rm nullity}(\mathcal{F}(\mathbb{D}))>1$) }}
We consider once again system \eqref{eq:example_1a}. This time the dictionary of functions $Z(x)$ is chosen as follows
\be\label{Z-extended}
\ba{l}
Z(x) =
{\rm col}(x,  
x^2,  
x^3, 
e^{x_1-x_2}-1,  
e^{x_1-x_2} x,
e^{x_1-x_2}x_1^2, 
\\
\hspace{1.8cm}
e^{x_1-x_2} x_1 x_2,  e^{x_1-x_2} x_1^3,
e^{x_1-x_2} x_2^2,
\\
\hspace{1.8cm}
e^{x_1-x_2} x_1^2 x_2, e^{x_1-x_2} x_1^4)
\ea\ee
where, as before, by $x^2$ we mean the vector with components $x_1^2$ and $x_2^2$, and $x^3, e^{x_1-x_2} x$ have a similar meaning. 
Moreover, $Y(x)=Z(x)$ and $W(x)={\rm col}(1, Z(x))$. 

We first explain the rationale behind the  extension of $Z(x)$  as in \eqref{Z-extended}. We have argued in Example \ref{ex:MB} that a function that returns the change of coordinates in the feedback linearization problem is the solution of  the partial differential equation 
$L_g \lambda (x) =0 $
with the nontriviality condition $L_{[f,g]}\lambda(x^0)\ne 0$. As $g(x)=\begin{bmatrix} 1 & 1 \end{bmatrix}$, the equation  $L_g \lambda (x) =0 $ is a special case of the transport equation, whose  general solution, discussed in Section \ref{sec:transport-eq} for the sake of completeness,   is $\lambda(x)=\zeta(x_1-x_2)$ where $\zeta$ is an arbitrary smooth function in a neighborhood of $x^0$. With  $Z(x)$ as in \eqref{Z-extended},  more  changes of coordinates are potentially obtainable than  from the choice of $Z(x)$  in Example \ref{ex:DD}, such as, for instance, $T_1 Z(x)= x_1-x_2 + {\rm e}^{x_1-x_2} (x_1-x_2)^2$, obtained by setting 
\[
T_1=\begin{bmatrix} 1 & -1 & 0_{1\times 7} & 1 & -2 & 0 & 1 & 0_{1\times 2}\end{bmatrix}.
\]

We also observe that the vector fields of the system can be expressed as $f(x)= A_* Z(x)$ and $g(x)= B_* W(x)$. Hence, as remarked in Subsection \ref{subsec:basis-functs}, we can determine the basis functions in $\phi(x,u)$ by selecting independent functions from $Z(x)$, $Y(x)$, $W(x)$, $W(x)u$, $Z(x)\otimes \frac{\partial Z}{\partial x}^\top$, $W(x)u\otimes \frac{\partial Z}{\partial x}^\top$, from which  
$\phi(x,u)$ is determined. 
We have run an experiment as in Example \ref{ex:DD} 
and we checked condition \eqref{data-rich-cond},  which turns out to be satisfied
. Hence, we can apply Theorem \ref{thm:DD-2} (ii) and determine the matrices $T,M,N$ from $\ker(\mathcal{F}(\mathbb{D}))$. In this example,
the  dimension of $\ker(\mathcal{F}(\mathbb{D}))$ is $\nu=2$ and the null space is spanned by an orthonormal basis $\{h_1, h_2\}$. From $h_1$, we compute a triplet $(T^{(1)},M^{(1)},N^{(1)})={\rm vec}^{-1}(h_1)$ 
\[
\ba{rl}
T^{(1)}=& h_{1,1} \begin{bmatrix}
1 & -1 & 0 & 0_{1\times 12}\\
\mu & -\lambda & \lambda & 0_{1\times 12}\\
\end{bmatrix}
+\\[4mm]
&
h_{1,13}
\begin{bmatrix}
0_{1\times 6} &  1 & 0 & 0 &0 & 0_{1\times 5}\\
0_{1\times 6} &  0 & \mu & -\lambda & \lambda & 0_{1\times 5}\\
\end{bmatrix}
\ea
\]
\[
\ba{l}
N^{(1)}= h_{1,1}\begin{bmatrix} \mu^2 & -\lambda^2 & \lambda^2+2\lambda\mu & 0_{1\times 12}\end{bmatrix}\\
+h_{1,13} 
[\;
0_{1\times 7}\; 
\mu^2\;
-\lambda^2\; 
(\lambda + \mu)^2\; 
- 2\lambda \mu\;
2 \lambda \mu\;
\lambda^2\;
- 2 \lambda^2\;
\lambda^2]
\ea
\]
and 
\[\ba{rl}
M^{(1)}= &h_{1,1}\begin{bmatrix} -\lambda+\mu & 2\lambda & 0 & 0_{1\times 13}\end{bmatrix}\\
&+h_{1,13} 
\begin{bmatrix} 
 -\lambda+\mu &
0_{1\times 6} & 
 -\lambda+\mu &
2\lambda & 
0_{1\times 7} 
\end{bmatrix}
\ea\]
where $h_{1,1}=0.5794$ and $h_{1,13}=-0.1079$ denote the entries $1$ and $13$ of $h_1\in \mathbb{R}^{15}$.
We also observe that the vector $h_1$ determines the output function 
\[
T_1^{(1)} Z(x)= h_{1,1}(x_1-x_2)+h_{1,13} (e^{x_1-x_2}-1)
\]
obtained from the linear combination of the functions in $Z(x)$.  

Similarly, from $h_2$, we compute a triplet $(T^{(2)},M^{2)},N^{(2)})={\rm vec}^{-1}(h_2)$
\[\ba{rl}
T^{(2)}=& 
h_{2,13}
\begin{bmatrix}
0_{1\times 6} &  1 & 0 & 0 &0 & 0_{1\times 5}\\
0_{1\times 6} &  0 & \mu & -\lambda & \lambda & 0_{1\times 5}\\
\end{bmatrix},
\ea
\]
which gives the output function 
\[
T_1^{(2)} Z(x)= h_{2,13} (e^{x_1-x_2}-1)
\]
where $h_{2,13}=0.6164$. The expressions of $M^{(2)}, N^{(2)}$ are omitted.

To understand the reason for the specific values of $h_1,h_2$ and ${\rm nullity}(\mathcal{F}(\mathbb{D}))=2$, we consider the 
conditions \eqref{cond-rel-degree}, \eqref{inv-cond-onT1} in Proposition \ref{lem:dimension-nu-space-V}, and characterize which structure these 
conditions impose on a generic function $T_1 Z(x)$, that is, on a function that is a linear combination of the functions in $Z(x)$ via the coefficients in $T_1=\begin{bmatrix} a_1 & \ldots, a_{15}\end{bmatrix}$. By Proposition \ref{lem:dimension-nu-space-V}, the specific functions $T_1 Z(x)$ determined by $h_1, h_2$ will have the same structure. 

In this example, the condition  \eqref{cond-rel-degree} simply becomes $T_{1} L_{g} Z(x)=0$. 
Recalling that $T_1=\begin{bmatrix} a_1 & \ldots, a_{15}\end{bmatrix}$, the condition \eqref{cond-rel-degree} imposes $a_2=-a_1$,  $a_9=-a_8$, $a_{11}=2a_{10}$, $a_{13}=a_{10}$, while the remaining entries of $T_1$ must be equal  to zero. 
Hence, this condition excludes  those functions that would make $T_1Z(x)$ not expressible as a smooth function of $x_1-x_2$ and returns 
\[
\ba{l}
T_1Z(x)=
a_1(x_1-x_2)+a_7 (e^{x_1-x_2}-1)\\[1mm]
\;\;\qquad\qquad+a_8 e^{x_1-x_2}(x_1-x_2)+a_{10}e^{x_1-x_2}(x_1-x_2)^2.
\ea\] 
One can further compute $T_1 L_f Z(x)$ and observe that the single term $a_{10}\lambda e^{x_1-x_2} x_1^2  x_2^2$ appears in it. Since $e^{x_1-x_2} x_1^2  x_2^2$ is not part of $Z(x)$, to have the first of the conditions \eqref{inv-cond-onT1} satisfied, i.e.,   $T_1 L_f Z(x)$ in ${\rm span}_{\mathbb{R}}\{Z_1(x), \ldots, Z_s(x)\}$, it must be $a_{10}=0$. Hence,
\[
\hspace{-4.5mm}
\ba{rl}
&L_f T_1Z(x)=T_1 L_f Z(x)=\\
&(\mu x_1+\lambda x_1^2 -\lambda x_2) [a_1 +(a_7+a_8)e^{y}+a_8e^{y}y]_{y=x_1-x_2}
\ea\]
Further calculations show that 
\[
\hspace{-4.5mm}
\ba{rl}
&L_f ^2T_1Z(x)=T_1 L_f^2 Z(x)=\\[2mm]
&(\mu x_1+\lambda x_1^2 -\lambda x_2)^2 [(a_7+2a_8)e^{y}+a_8e^{y}y]_{y=x_1-x_2}\\[2mm]
&+(\mu^2 x_1+(2\mu\lambda+\lambda^2) x_1^2 -\lambda^2 x_2)
 [a_1 +(a_7+a_8)e^{y}\\[2mm]
 &+a_8e^{y}y]_{y=x_1-x_2}
\ea\]
In this expression the single term $a_8\lambda^2 e^{x_1-x_2} x_1^5$ appears, which imposes $a_8=0$. In doing so, we obtain that $T_1 L_f^2 Z(x)$ belongs to ${\rm span}_{\mathbb{R}}
\{Y_1(x), \ldots, Y_p(x)\}$ (recall that, for this example, ${\rm span}_{\mathbb{R}}
\{Y_1(x), \ldots, Y_p(x)\}={\rm span}_{\mathbb{R}}
\{Z_1(x), \ldots, Z_s(x)\}$), that is, the second condition in \eqref{inv-cond-onT1} is satisfied. Then, one  also determines the matrix $N$ such that $T_1 L_f^2 Z(x) = N Y(x)$, 
which gives 
\[\ba{l}
N=[\;a_1 \mu^2\; 
- a_1 \lambda^2\;  
a_1 (\lambda^2 +2 \lambda \mu)\; 
0_{1\times 4}\; 
a_7 \mu^2\;
- a_7 \lambda^2\; \\
\quad 
a_7 (\lambda + \mu)^2\; 
- 2 a_7 \lambda \mu\;
2 a_7 \lambda \mu\;
a_7 \lambda^2\;
- 2 a_7 \lambda^2\;
a_7 \lambda^2]
\ea
\]
Finally, we have
\[
\ba{rl}
&L_g L_f T_1Z(x)=T_1 L_g L_f Z(x)=a_1 (-\lambda+ \mu) \\[1mm]
&+ 2 a_1 \lambda x_1  +a_7 (-\lambda+\mu)  e^{x_1 - x_2} + 
 2 a_7 \lambda e^{x_1 - x_2}  x_1
\ea\]
Since we chose $W(x)=\begin{bmatrix}1&Z(x)^\top \end{bmatrix}^\top$ as in Example \ref{ex:DD}, we see that $T_1 L_g L_f Z(x)$ belongs to ${\rm span}_{\mathbb{R}}\{W_{1}(x), \ldots, W_{r}(x)\}={\rm span}_{\mathbb{R}}\{1, Z_1(x), \ldots, Z_s(x)\}$, i.e., the last condition of  \eqref{inv-cond-onT1} holds, and the row vector  $M$ in the identity $T_1 L_g L_f Z(x)= MW(x)$ is given by 
\[
M\!\!=\!\![(a_1+ a_7) (-\lambda+ \mu)\; 2 a_1 \lambda\; 0_{1\times 5}\; a_7 (-\lambda+\mu)\;
2 a_7\lambda\; 0_{1\times 7}].
\]
In conclusion, 
the conditions \eqref{cond-rel-degree}, \eqref{inv-cond-onT1} impose that $T_1$ has only two nonzero entries, $a_1$, $a_7$, therefore
\[
T_2= 
\begin{bmatrix}
a_1 \mu &-a_1 \lambda & a_1 \lambda & 0_{1\times 4} &a_7 \mu
& -a_7 \lambda& a_7 \lambda & 0_{1\times 5} 
\end{bmatrix}.
\] 
This implies   that the resulting function $T_1 Z(x)$ is 
\[
\ba{l}
T_1Z(x)=
a_1(x_1-x_2)+a_7 (e^{x_1-x_2}-1).
\ea\] 
We obtain the specific function $T_1^{(1)}Z(x)$ determined by $h_1$ by setting $a_1=h_{1,1}$ and $a_7=h_{1,13}$ and $T_1^{(2)}Z(x)$ determined by $h_2$ by setting $a_1=0$ and $a_7=h_{1,13}$. Consequently, for these values of $T_1^{(1)},T_1^{(2)}$ we also obtain the functions $N^{(1)}Y(x)$, $M^{(1)}W(x)$ and $N^{(2)}Y(x)$, $M^{(2)}W(x)$.  \qedp
\end{example}

\medskip

According to Theorem \ref{thm:DD-2}, under Assumption \ref{asspt:data-rich} on the dataset being sufficiently rich, we can determine the functions $\tau(x), \gamma(x), \delta(x)$ that satisfy \eqref{syst.after.change.coord} by computing the null space of the matrix $\mathcal{F}(\mathbb{D})$. Condition $v\in \ker(\mathcal{F}(\mathbb{D}))$, however, does not encode any requirement on the resulting matrices $(T, M, N)={\rm vec}^{-1}(v)$ to enforce that $TZ(x)$ is a change of coordinates and $MW(x)$ is a nonsingular function in a neighbourhood of $x^0$. In fact, Proposition \ref{lem:dimension-nu-space-V} runs short of proving that these properties hold for multi-input systems and Example \ref{ex:MIMO} show that  in general they are not guaranteed under the given assumptions. Nonetheless, in the case of single input systems, any nonzero $v$  for which $\mathcal{F}(\mathbb{D})v=0$ returns matrices $T,M$ with such desired properties.
This is a consequence of Assumption \ref{ass:lin-independence} and the following ensuing result. In view of this result and Proposition \ref{lem:dimension-nu-space-V}, we conclude that, for single input systems, by solving $\mathcal{F}(\mathbb{D})v=0$, we determine output functions that are expressible through linear combinations of the functions in $Z(x)$ and with respect to which the system \eqref{eq:system} has relative degree $n$ at $x^0$. 
\begin{proposition}\label{lem:1-dimension}
Let $m=1$. Let Assumptions \ref{ass:elp-solvable}--\ref{ass:lin-independence} hold
and $g(x^0)\ne 0$. 
Consider any nonzero vector $v$ that satisfies $\mathcal{F}(\mathbb{D})v=0$ and the associated matrices $(T, M, N)={\rm vec}^{-1}(v)$.  Then $TZ(x)$ is a change of coordinates and $MW(x)$ is a nonzero function in a neighbourhood of $x^0$. \qeds
\end{proposition}

\emph{Proof.} See Section \ref{sec:proof-prop-4}. \qedp

\medskip

The following example shows that the last result does not carry over to multi-input systems. 

\begin{example} \label{ex:MIMO}
Consider the MIMO nonlinear control system \cite[Example 5.2.6]{isidori1995book}
\begin{equation}\label{mimo.sys}
\dot x= \begin{bmatrix}
x_2 + x_2^2\\
x_3 -x_1 x_4 + x_4 x_5\\
x_2 x_4 + x_1 x_5 - x_5^2\\ 
x_5\\
x_2^2
\end{bmatrix}
+
\begin{bmatrix}
0\\
0\\
\cos(x_1-x_5) \\
0\\
0
\end{bmatrix}u_1
+
\begin{bmatrix}
1\\
0\\
1\\
0\\
1
\end{bmatrix} u_2 
\end{equation}

In \cite[Example 5.2.6]{isidori1995book} it is shown that the system is feedback linearizable  at $x^0=0$ 
 (Assumption \ref{ass:elp-solvable} holds), with $\mathcal{D}$ any neighborhood  of $x^0=0$ contained in the set $\{x\in \mathbb{R}^5\colon |x_1-x_5|<\frac{\pi}{2}\}$.   Assumption \ref{ass:library} holds with 
\[
Z(x)= \begin{bmatrix}
x\\
x_1 x_4\\
x_4 x_5\\
x_2^2
\end{bmatrix}=Y(x), \; W(x)=\begin{bmatrix}
1 & 0\\
0 & 1 \\
c_{15}& 0\\
0 & c_{15}
\end{bmatrix}. 
\]
where $c_{15}:=\cos(x_1-x_5)$,  and 
\[
A_c = \begin{bmatrix}
0 & 1 & 0 & 0 & 0 \\
0 & 0 & 1 & 0 & 0 \\
0 & 0 & 0 & 0 & 0 \\
0 & 0 & 0 & 0 & 1 \\
0 & 0 & 0 & 0 & 0
\end{bmatrix}, 
B_c = \begin{bmatrix}
0 & 0\\
0 & 0\\
1 & 0\\
0 & 0 \\
0 & 1
\end{bmatrix}.
\]
An example of matrices $\overline T,\overline M,\overline N$ that satisfy Assumption \ref{ass:library} is 
\be\label{vbar-exmpl-MIMO}
\begin{array}{l}
\overline T = 
\begin{bmatrix}
1 & 0 & 0 & 0 & -1 & 0 & 0& 0\\
0 & 1 & 0 & 0 & 0 & 0 & 0& 0\\
0 & 0 & 1 & 0 & 0 & -1 & 1& 0\\
0 & 0 & 0 & 1 & 0 & 0 & 0& 0\\
0 & 0 & 0 & 0 & 1 & 0 & 0& 0\\
\end{bmatrix},\\[10mm]
\overline N = 
\begin{bmatrix}
0 & 0 & 0 & 0 & 0 & 0 & 0& 0\\
0 & 0 & 0 & 0 & 0 & 0 & 0& 1\\
\end{bmatrix},\\[4mm]
\overline M = 
\begin{bmatrix}
0 & 1 & 1 & 0\\
0 & 1 & 0 & 0\\
\end{bmatrix}.
\end{array}
\ee

Based on $Z(x), \frac{\partial Z}{\partial x}, Y(x), W(x)$ we construct the matrix $F(x,u,\dot x)$.  We asses it at the sample points of the dataset to obtain the matrix $\mathcal{F}(\mathbb{D})$. We determine that the null space of $\mathcal{F}(\mathbb{D})$ has dimension $4$, i.e., $\text{nullity}(\mathcal{F}(\mathbb{D}))=4>1$. We note that $\mathcal{F}(\mathbb{D})\overline v=0$, where $\overline v = \left[\begin{smallmatrix}
\text{vec}(\overline T)\\ \text{vec}(\overline M) \\ \text{vec}(\overline N)
\end{smallmatrix}\right]$, that is, the matrices \eqref{vbar-exmpl-MIMO} are among the solutions $v$ that we compute solving $\mathcal{F}(\mathbb{D})\overline v=0$. This is consistent with the theory: since $\overline v$ satisfies  $F(x,u,\dot x)\overline v=0$ for all $x,u\in \mathcal{D}\times \mathbb{R}^m$ and all $\dot x=f(x)+g(x)u$, then the same holds for $(x,u,\dot x)$ at the samples, which gives $\mathcal{F}(\mathbb{D})\overline v=0$. 

We verify that for the given dataset condition \eqref{data-rich-cond} holds.
Then Theorem \ref{thm:DD-2} can be applied and we expect that  any $v$ that belongs to $\ker(\mathcal{F}(\mathbb{D}))$ is a solution to $F(x,u,\dot x) v=0$. This can be numerically confirmed by considering  a basis $v_1, v_2, v_3, v_4$ of $\ker(\mathcal{F}(\mathbb{D}))$ and checking that $F(x,u,\dot x)v_i=0$ for $i=1,\ldots, 4$. In fact,  for any $v \in \ker(\mathcal{F}(\mathbb{D}))$, there exists coefficients $\alpha_i\in \mathbb{R}$, $i=1,\ldots,4$, such that $v=\sum_{i=1}^4 \alpha_i v_i$, from which we trivially have that $F(x,u,\dot x)v= \sum_{i=1}^4 \alpha_i F(x,u,\dot x)v_i =0$.

As an example,  we pick the first vector $v_1$ of a (orthonormal)  basis,  and we construct the matrices $T^{(1)}, M^{(1)}, N^{(1)}$ below\footnote{The matrices corresponding to the other vectors of the basis are given in Section \ref{subsec:details-example-MIMO}.} 
\[\begin{array}{l}
T^{(1)} = -0.3536
\begin{bmatrix}
1 & 0 & 0 & 0 & -1 & 0 & 0& 0\\
0 & 1 & 0 & 0 & 0 & 0 & 0& 0\\
0 & 0 & 1 & 0 & 0 & -1 & 1& 0\\
0 & 0 & 0 & 0 & 0 & 0 & 0& 0\\
0 & 0 & 0 & 0 & 0 & 0 & 0& 0\\
\end{bmatrix}, \\
N^{(1)} = 
\begin{bmatrix}
0 & 0 & 0 & 0 & 0 & 0 & 0& 0\\
0 & 0 & 0 & 0 & 0 & 0 & 0& 0\\
\end{bmatrix},\\
M^{(1)}= -0.3536
\begin{bmatrix}
0 & 1 & 1 & 0\\
0 & 0 & 0 & 0\\
\end{bmatrix}
\end{array}
\]
It is easily verified that this is indeed a solution to $F(x,u,\dot x)v=0$, but it is not a solution of interest because $T^{(1)}Z(x)$ is not a coordinate transformation and $M^{(1)} W(x)$ is a singular matrix. This does not contradicts our theoretical findings since our conditions do not encode any requirement on the nonsingularity of the Jacobian of $TZ(x)$ and $W(x)$. In fact, it is consistent with Proposition \ref{lem:dimension-nu-space-V}. The finding highlights the interesting point that, except for single input systems (for which Proposition \ref{lem:1-dimension} holds), even under the condition \eqref{data-rich-cond}, even though all the vectors in $\ker(\mathcal{F}(\mathbb{D}))$ satisfy the condition $F(x,u,\dot x)v=0$, among them we need to pick one $v = \left[\begin{smallmatrix}
\text{vec}(T)\\ \text{vec}(M) \\ \text{vec}(N)
\end{smallmatrix}\right]$ for which the resulting $T Z(x)$ is a coordinate transformation and $M W(x)$ is a nonsingular matrix. Numerically this can be done by expressing $T,M$ as a linear combination of the basis of $\ker(\mathcal{F}(\mathbb{D}))$ and then determining the coefficients of the linear combination that make the matrices $T \frac{\partial Z}{\partial x}(x^0)$ and $M W(x^0)$ nonsingular. Note that these coefficients exist, since $\overline v \in \ker(\mathcal{F}(\mathbb{D}))$. They can be determined by expressing $\overline v = \left[\begin{smallmatrix}
\text{vec}(\overline T)\\ \text{vec}(\overline M) \\ \text{vec}(\overline N)
\end{smallmatrix}\right]$, where $\overline T,\overline M,\overline N$  are as in \eqref{vbar-exmpl-MIMO}, via the basis of $\ker(\mathcal{F}(\mathbb{D}))$, which in this case gives $\overline v= -2.824 v_1 + 1.8095 v_2 - 0.0011 v_3 - 0.8518 v_4$ (these coefficients are obtained by solving the equation $H x = \overline v$, where $H$ is the matrix whose columns are the vectors of the basis of    $\ker(\mathcal{F}(\mathbb{D}))$). Then $(\overline T,\overline M,\overline N)= 
-2.824 {\rm vec}^{-1}(v_1) + 1.8095 {\rm vec}^{-1}(v_2) - 0.0011 {\rm vec}^{-1}(v_3) - 0.8518 {\rm vec}^{-1}(v_4)$.  \qedp

\end{example}

\section{Concluding remarks} \label{sec:discussion}


In this paper, we have revisited the problem of feedback linearization through the lens of data. Starting from the ideal case where we have exact knowledge of the model, we have derived a characterization of the solutions in terms of (nonlinear) algebraic equations. We have then shown how this characterization lends itself well to be used in the context of data-driven control. In essence, we have shown how the problem of feedback linearization can be solved from data through the computation of the zeros of data-dependent 
nonlinear algebraic equations. To our knowledge, this is the first result that 
demonstrates how one can generalize from the sample space to the entire space of solutions, meaning how the solution represents a valid change of coordinates for feedback linearization even for states that were not considered in determining the solution. We have also discussed some important differences between the MIMO and SISO cases, the latter being the case where the solution can be directly found as the null space solution of certain data matrices.

The results presented in this paper represent a first step in understanding the problem of feedback linearization from data. Extensions of these ideas certainly include e.g. the input-output  
and the dynamic feedback \cite{charlet1989dynamic}, \cite[Section 5.4]{isidori1995book} linearization  problem
and, more generally, the immersion problem \cite{fliess2005finite}, which deals with linearization in spaces of dimensions larger than that of the original coordinate space. Other interesting extensions concern practical aspects typical of the data-driven control. Among these is certainly the study of the sensitivity (robustness) of the proposed approach to the case of noisy data. In this regard, a possible approach is to filter the data through averaging techniques \cite[Section VI.C]{NonlinearityCancellation2023} or to use integration techniques to filter out noise, as also highlighted in the context of data-driven control \cite[Appendix A]{cdp-rp-pt2023event}.


\section{appendix}

\begin{lem}\label{lem:renk-equivalence}
The matrix 
\be\label{phi.miniscule}
\begin{bmatrix}
\phi(x_0,u_0)^\top\\
\vdots\\
\phi(x_{L-1},u_{L-1})^\top\\
\end{bmatrix}
\ee
has full column rank if and only if the matrix
\be\label{phi.capital}
\begin{bmatrix}
\Phi(x_0, u_0)\\
\vdots\\
\Phi(x_{L-1}, u_{L-1})
\end{bmatrix}
\ee
has full column rank. \qeds
\end{lem}

\textit{Proof.} (Only if) Suppose by contradiction that there exists $w\in \mathbb{R}^{n n_b}\setminus\{0\}$ such that 
\[
\begin{bmatrix}
\Phi(x_0, u_0)\\
\ldots\\
\Phi(x_{L-1}, u_{L-1})
\end{bmatrix}w=0
\]
or, equivalently, such that 
\[
\Phi(x_{i}, u_{i})w=0, \forall i\in \{0,1, \ldots, L-1\}.
\]
By the definition $\Phi(x,u)=I_n \otimes \phi(x,u)^\top$, and partitioning $w$ as $\begin{bmatrix}w_1^\top & w_2^\top &\ldots & w_n^\top \end{bmatrix}$, where each sub-vector $w_k$ is in $\mathbb{R}^{n_b}$,  the identity above is equivalent to 
\[
\phi(x_{i}, u_{i})^\top w_j=0, \forall i\in \{0,1, \ldots, L-1\}, \forall j\in \{1,2, \ldots, n\}.
\]
Pick an index $j$ such that $w_j\ne 0$. Such an index exists because $w\ne 0$. Then, the last identity implies that 
\[
\phi(x_{i}, u_{i})^\top w_j=0, \forall i\in \{0,1, \ldots, L-1\}
\]
which contradicts the assumption that the matrix \eqref{phi.miniscule} has full column rank.

(If) Suppose by contradiction that \eqref{phi.miniscule} has not full column rank. Then there exists $0 \ne v\in \mathbb{R}^{n_b}$ such that  
\[
\phi(x_{i}, u_{i})^\top v=0, \forall i\in \{0,1, \ldots, L-1\}.
\]
By the definition $\Phi(x,u)=I_n \otimes \phi(x,u)^\top$,  the vector $w:=\mathds{1}_n\otimes v\ne 0$ satisfies 
\[
\Phi(x_{i}, u_{i}) w=0, \forall i\in \{0,1, \ldots, L-1\},
\]
which implies that the matrix \eqref{phi.capital} is not full column, which is a contradiction. This ends the proof. \qedp

\subsection{Proof of Proposition \ref{lem:dimension-nu-space-V}}\label{sec:proof-prop-2}

First we observe that the set of vectors $v$ that satisfy $\mathcal{F}(\mathbb{D})v=0$ is a subset of the vector space $\mathbb{R}^\mu$. This subset is a nonempty set  that is closed under the operation of vector addition and multiplication by a scalar, is a subspace of $\mathbb{R}^\mu$, hence a vector space. By Assumptions \ref{ass:elp-solvable} and \ref{ass:library} there exists a nonzero vectors $v$ that satisfy $\mathcal{F}(\mathbb{D})v=0$. Hence, the dimension of the space of solutions of $\mathcal{F}(\mathbb{D})v=0$ is greater than or equal to $1$. Any of these nonzero vectors $v$ that satisfy $\mathcal{F}(\mathbb{D})v=0$ also satisfies $F(x,u,\dot x)v=0$. 
By  \eqref{eq:MB_solution}, the triplet of matrices $(T,N,M)={\rm vec}^{-1}(v)$ satisfies 
\be\label{eq:MB_solution-aux1_MIMO}
T \displaystyle\frac{\partial Z}{\partial x} (f(x)+g(x)u)= 
A_c   T Z(x) + B_c(  N Y(x)+  M W(x)u),
\ee
for all $x\in \mathcal{D}$, all $u\in \mathbb{R}^m$. 
From the identity \eqref{eq:MB_solution-aux1_MIMO}, one deduces that, for every $i=1, 2, \ldots, m$,  
the function $T^i_1 Z(x)$
satisfy 
\be\label{chain-identities-mimo}
\hspace{-0.38cm}
\ba{rlrl}
T^i_1\frac{\partial Z}{\partial x} f(x)=& \!\!\!\!\!T^i_{2} Z(x) &  
T^i_1\frac{\partial Z}{\partial x} g(x)=& \!\!\!\!\! 0\\[2mm]
T^i_2\frac{\partial Z}{\partial x} f(x)=& \!\!\!\!\!T^i_{3} Z(x) &  
T^i_2\frac{\partial Z}{\partial x} g(x)=& \!\!\!\!\! 0\\
\vdots & & \vdots & \!\!\!\!\!\\
T^i_{r_i-1}\frac{\partial Z}{\partial x} f(x)=& \!\!\!\!\!T^i_{r_i} Z(x) &  
T^i_{r_i-1}\frac{\partial Z}{\partial x} g(x)=& \!\!\!\!\! 0\\
T^i_{r_i}\frac{\partial Z}{\partial x} f(x)=& \!\!\!\!\! N_i Y(x) &  
T^i_{r_i }\frac{\partial Z}{\partial x} g(x)=& \!\!\!\!\! M_i W(x)
\ea
\ee
for all $x\in \mathcal{D}$, all $u\in \mathbb{R}^m$.  
Thanks to this chain of identities, there exists  $i\in \{1, 2, \ldots, m\}$,  such that $T^i_1\ne 0$. In fact, suppose by contradiction that $T^i_1= 0$ for every $i\in \{1, 2, \ldots, m\}$. This implies $T^i_2 Z(x) =0$, which gives $T^i_2=0$ by Assumption \ref{ass:lin-independence}. Applying this argument recursively, we obtain that $T^i_1= 0$ implies $T^i_2=T^i_3=\ldots= T^i_{r_i}=0$. This would also imply that $N_i Y(x)=0$ and $M_i W(x) = 0$, which would return $N_i=0$ and $M_i=0$ by Assumption \ref{ass:lin-independence}. As this holds for every $i\in \{1, 2, \ldots, m\}$, we would conclude that $T=0$, $N=0$ and $M=0$, which contradicts $v\ne 0$. Hence, there exists $i\in \{1, 2, \ldots, m\}$ such that  $T^i_1\ne 0$, and the function $T^i_1 Z(x)$ is not an identically zero function.  The chain of identities in \eqref{chain-identities-mimo} also shows that
\be\label{chain-identities-mimo-consequence}
\ba{rl}
 T^i_{1} L_f Z(x)=
 & T^i_{2} Z(x)\\ 
T^i_{1}  L_f^2 Z(x) =
& T^i_{3} Z(x)\\
\vdots& \\
 T^i_{1}  L_f^{r_i-1}Z(x) =& T^i_{r_i} Z(x).
\ea\ee
These identities have several implications. 
The first one is that  the functions  $T^i_{1} L_f Z(x), T^i_{1}  L_f^2 Z(x),  \ldots, T^i_{1}  L_f^{r_i-1}Z(x)$, or, equivalently,  $L_f T^i_{1}   Z(x), L_f^2 T^i_{1}   Z(x),  \ldots, L_f^{r_i-1} T^i_{1}  Z(x)$, must belong to ${\rm span}_{\mathbb{R}}\{Z_1(x), \ldots, Z_s(x)\}$.  The second one is that  the vectors $T^i_{2},  \ldots, T^i_{r_i}$ linearly depend on $T^i_{1}$. In fact, the function $L_f^j Z(x)$ on left-hand side of the identity $T^i_{1} L_f^j Z(x)=T^i_{j+1} Z(x)$ can be written as $\Xi^j_a Z(x)+\Xi^j_b \hat Z^j(x)$, where $\hat Z^j(x)$ is a vector that collects all the functions appearing in $L_f^j Z(x)$ not part of $Z(x)$, and $\Xi^j_a, \Xi^j_b$ are matrices of coefficients that multiply the functions in $Z(x)$ and $\hat Z^j(x)$ appearing in $L_f^j Z(x)$. Hence  the identity $T^i_{1} L_f^j Z(x)=T^i_{j+1} Z(x)$ implies that $T^i_{j+1}=T^i_{1}\Xi^j_a$ and $0=T^i_{1}\Xi^j_b$. 

In addition, the identity $T^i_{r_i} L_f Z(x)= N_i Y(x)$ is equivalent to $T^i_{1}  L_f^{r_i}Z(x) = N_i Y(x)$, which shows that $T^i_{1}  L_f^{r_i}Z(x)$ belongs to ${\rm span}_{\mathbb{R}}\{Y_1(x), \ldots, Y_p(x)\}$, where the functions $Y_j(x)$, $j=1, \ldots, p$, are the entries of $Y(x)$. With the same arguments as above, $T^i_{1}  L_f^{r_i}Z(x) = N_i Y(x)$ also points out that $N_i$ linearly depends on $T^i_{1}$. Similarly, one shows that, for every $1\le i,k\le m$,  $T^i_{1}  L_{g_k} L_f^{r_i-1}Z(x) = M_i W_k(x)$,  where $W_k(x)$ is the $k$th column of $W(x)$, leading to the conclusion  that  $T^i_{1}  L_{g_k} L_f^{r_i-1}Z(x)$ belongs to ${\rm span}_{\mathbb{R}}\{W_{1k}(x), \ldots, W_{rk}(x)\}$ and that $M_i$ linearly depend on $T^i_{1}$. 
The equations in the left column of \eqref{chain-identities-mimo}
further show that,  for every $1\le i,k\le m$ and every $0\le j\le r_i-2$,  $T^i_{1} L_{g_k} L_f^{j}Z(x)=0$. \qedp

The proof specialized to the case of single input systems (Section \ref{sec:proof-prop-4}) shows that the solution $(T,N,M)={\rm vec}^{-1}(v)$ is such that $TZ(x)$ is a change of coordinates and $MW(x)$ is nonsingular. The arguments used for the single input case cannot be extended to the multi input case and Example \ref{ex:MIMO} shows that the result does not hold under the current assumptions. By \cite[Lemma 5.1.1]{isidori1995book}, for multi input systems we can conclude the weaker result that, if the matrix $M$ is such that $M W(x)$ is nonsingular, then $TZ(x)$ is a change of coordinates.

\subsection{Proof of Proposition \ref{prop3}}\label{sec:proof-prop3}
It is straightforward and therefore just sketched. By Assumptions \ref{ass:elp-solvable} and \ref{ass:library}, equation \eqref{eq:MB_solution} holds, in which we replace $\overline T,\overline  M,\overline  N$ with $T,M,N$. This implies that $F(x,u, \dot x)v=0$ holds with 
$v:={\rm vec}(T,N,M)$. By Theorem \ref{thm:DD-2} (i),  $\mathcal{F}(\mathbb{D})v=0$.
Equation \eqref{eq:MB_solution} also shows that for every $i\in \{1,\ldots,m\}$ there exist row vectors $T^i_1\ne 0$ that satisfy \eqref{cond-rel-degree}, \eqref{inv-cond-onT1}. The first condition in \eqref{cond-rel-degree} implies that the row vectors $T^i_2,\ldots, T^i_{r_i}$ are such that $T^i_{1}  L_f^{k}Z(x)= T^i_{k+1} Z(x)$, for $1\le k\le  r_i-1$, thus showing their dependence on $T^i_{1}$. The other conditions in \eqref{inv-cond-onT1} show the dependence of  $N_i,M_i$ on $T^i_1$. \qedp

\subsection{Further details on Example \ref{empl:extended-Z}}\label{subsec:details-example-extZ}

\subsubsection{General solution to $L_g\lambda(x)=0$}\label{sec:transport-eq}

The general solution to the (special case of the) 
transport equation   $L_g\lambda(x)=0$, or $\frac{\partial \lambda}{\partial x_1}+\frac{\partial \lambda}{\partial x_2}=0$, is usually determined by the method of characteristics. Here, we consider a derivation based on a change of variables (see, e.g., \cite{novozhilov}). Consider the linear change of variables $z_1=x_1$ and $z_2=x_2-x_1$ and let $\tilde\lambda(z):=[\lambda(x)]_{x_1=z_1, 
x_2=z_1+z_2}$, or, by the invertibility of the change of variables,
$[\tilde\lambda(z)]_{z_1=x_1, z_2=x_2-x_1}=\lambda(x)$.
Then 
\[\ba{l}
\displaystyle\frac{\partial  \lambda}{\partial x_1}= \left[\frac{\partial \tilde\lambda}{\partial z_1}\right]_{{\scriptsize\!\!\!\!\ba{l}z_1=x_1,\\ z_2=x_2-x_1\ea}}-\left[\frac{\partial \tilde\lambda}{\partial z_2}\right]_{{\scriptsize\!\!\!\!\ba{l}z_1=x_1,\\ z_2=x_2-x_1\ea}}
\\
\displaystyle\frac{\partial  \lambda}{\partial x_2}=\left[\frac{\partial \tilde\lambda}{\partial z_2}\right]_{{\scriptsize\!\!\!\!\ba{l}z_1=x_1,\\ z_2=x_2-x_1\ea}},
\ea\]
which imply 
$\left[\frac{\partial \tilde\lambda}{\partial z_1}\right]_{{\tiny\!\!\!\!\ba{l}z_1=x_1,\\ z_2=x_2-x_1\ea}}=0$.
Hence, $\tilde\lambda(z)=\zeta(z_2)$, where $\zeta$ is an arbitrary smooth function. In the original coordinates, this gives $\lambda(x)=\zeta(x_2-x_1)$.

\subsubsection{Details on the computation of $L_g T_1 Z(x)=0$}\label{sec:transport-equation}
The constraint $L_g T_1 Z(x)=0$ imposes that  $\frac{\partial T_1 Z(x)}{\partial x_1}+\frac{\partial T_1 Z(x)}{\partial x_2}=0$. 
We first compute the two entries of the gradient  of $T_1 Z(x)$.
\[\ba{l}\frac{\partial T_1 Z(x)}{\partial x_1}=\\
a_1  + 2 a_3 x_1+ 3 a_5 x_1^2 \\
+ (a_7+a_8) e^{x_1 - x_2} \\
 + (a_8 + 2 a_{10}) e^{x_1 - x_2} x_1+ (a_9 +a_{11}) e^{x_1 - x_2} x_2 \\
  + (a_{10} + 3 a_{12}) e^{x_1 - x_2} x_1^2    \\
+ (a_{11} + 2 a_{14}) e^{x_1 - x_2} x_{1} x_{2}  \\
+ (a_{12} + 4 a_{15}) e^{x_1 - x_2} x_{1}^3\\ 
+ 
 a_{13} e^{x_1 - x_2} x_2^2\\
+ a_{14} e^{x_1 - x_2} x_1^2 x_2 \\
  + a_{15} e^{x_1 - x_2} x_1^4  \\
\ea \]
and
\[\ba{l}
\frac{\partial T_1 Z(x)}{\partial x_2}=\\
a_2 + 
 2 a_4 x_2+ 3 a_6 x_2^2  \\
+(- a_7+ a_9) e^{x_1 - x_2}\\  
+( - a_8+ a_{11}) e^{x_1 - x_2} x_1 +(- a_9 + 2 a_{13}) e^{x_1 - x_2} x_2 \\ 
+( - a_{10} + a_{14}) e^{x_1 - x_2} x_1^2 \\
 - a_{11} e^{x_1 - x_2} x_1 x_2 \\
 - a_{12} e^{x_1 - x_2} x_1^3 \\
  - a_{13} e^{x_1 - x_2} x_2^2\\
 - a_{14} e^{x_1 - x_2} x_1^2 x_2\\ 
 - a_{15} e^{x_1 - x_2} x_1^4. 
\ea\]
We observe that all the  functions appearing in  $\frac{\partial T_1 Z(x)}{\partial x_1}+\frac{\partial T_1 Z(x)}{\partial x_2}$ are linearly independent, thus for the equation to hold the coefficients that multiply such independent functions must be zero. Hence, it holds that
\[
\ba{l}
a_1+a_2=0, \;a_3=0, \; a_4=0, \; a_5=0,\; a_6=0,\\
a_8+a_9=0\\
2a_{10}+a_{11}=0,\; a_{11}+2a_{13}=0\\
3 a_{12}+a_{14}=0\\
a_{14}=0\\
a_{15}=0\\
\ea
\]
from which
\[
\ba{l}
a_2=-a_1, \;a_3=0, \; a_4=0, \; a_5=0,\; a_6=0,\\
a_9=-a_8\\
a_{11}=2a_{10},\; a_{13}=a_{10}\\
a_{12}=0\\
a_{14}=0\\
a_{15}=0\\
\ea
\]
Thanks to this constraint, $T_1Z(x)$ reduces to 
\[
\ba{l}
T_1Z(x)=
a_1(x_1-x_2)+a_7 (e^{x_1-x_2}-1)\\[1mm]
\;\;\qquad\qquad+a_8 e^{x_1-x_2}(x_1-x_2)+a_{10}e^{x_1-x_2}(x_1-x_2)^2.
\ea\]

\subsection{Proof of Proposition \ref{lem:1-dimension}}\label{sec:proof-prop-4}

We specialize the arguments of the proof of Proposition \ref{lem:dimension-nu-space-V} up to formula \eqref{chain-identities-mimo-consequence} to the case $m=1$. These arguments show that 
$T_1\ne 0$ and  $T_1Z(x)$ is not an identically zero function by Assumption \ref{ass:lin-independence}. 

Recall now that Assumption \ref{ass:elp-solvable} and $\text{rank}(g(x^0))=m$ imply that condition \eqref{syst.after.change.coord} holds. That is, there exists a function $\tau_1(x)$ such that the single input system \eqref{eq:system} with the  output function  $\tau_1(x)$ has relative degree $n$. Following the proof of 
\cite[Lemma 4.1.1 and Lemma 4.2.2]{isidori1995book}, a number of properties can be established as a consequence of the existence of a relative degree $n$ with respect to some output function. First, the matrix 
$\left[\begin{smallmatrix} 
g(x^0) & \text{ad}_f g(x^0)& \ldots & \text{ad}_f^{n-1} g(x^0)
\end{smallmatrix}\right]$ has rank $n$. Hence, the distribution 
$D:={\rm span}\{g, {\rm ad}_f g, \ldots, {\rm ad}_f^{n-2} g\}$ is nonsingular, because its dimension is $n-1$ in a neighborhood of $x^0$. We also know (\cite[p.~22]{isidori1995book}) that  $D^\perp$ is a smooth, nonsingular codistribution, has dimension $1$ and around $x^0$ it is spanned by a covector field.  

We turn our attention again to the function $T_1 Z(x)$. 
By the chain of the identities 
\eqref{chain-identities-mimo} specialized to the case $m=1$, 
we know that 
$\frac{\partial T_1  Z(x)}{\partial x}
\left[\begin{smallmatrix} 
g(x) & \text{ad}_f g(x)& \ldots & \text{ad}_f^{n-2} g(x)
\end{smallmatrix}\right]=0$. 
Since we have already established that $T_1  Z(x)$ is not an identically zero function and that 
$D^\perp=({\rm span}\{g, {\rm ad}_f g, \ldots, {\rm ad}_f^{n-2} g\})^\perp$ is a smooth nonsingular codistribution of dimension $1$ around $x^0$, the last equality above allows us to conclude that $\frac{\partial T_1  Z(x)}{\partial x}$ is a covector field spanning $D^\perp$, that is, a basis of the  
$1$-dimensional codistribution $D^\perp$. Additionally, we observe that $\frac{\partial T_1  Z}{\partial x}\text{ad}_f^{n-1} g(x^0)\ne 0$, otherwise the identity  $\frac{\partial T_1  Z(x)}{\partial x}
\left[\begin{smallmatrix} 
g(x) & \text{ad}_f g(x)& \ldots & \text{ad}_f^{n-2} g(x)
\end{smallmatrix}\right]=0$ and the non-singularity of the matrix $\left[\begin{smallmatrix} 
g(x^0) & \text{ad}_f g(x^0)& \ldots & \text{ad}_f^{n-1} g(x^0)
\end{smallmatrix}\right]$ would imply that $\frac{\partial T_1  Z}{\partial x}(x^0)= 0$, which contradicts that $\frac{\partial T_1  Z}{\partial x}(x)$ is the basis of a $1$-dimensional codistribution in a neighborhood of $x^0$ (hence, its dimension must be constant and equal to $1$ in such a neighborhood). 

Having shown that the function $T_1 Z(x)$ satisfies $\frac{\partial T_1 Z (x)}{\partial x}
\left[\begin{smallmatrix} 
g(x) & \text{ad}_f g(x)& \ldots & \text{ad}_f^{n-2} g(x)
\end{smallmatrix}\right]=0$ in a neighborhood of $x^0$ and the nontriviality condition 
$\frac{\partial T_1 Z(x)}{\partial x}\text{ad}_f^{n-1} g(x^0)\ne 0$, we can follow the proof of \cite[Lemma 4.1.1]{isidori1995book} to conclude that $\frac{\partial T_1  Z}{\partial x}(x^0), \frac{\partial L_f T_1 Z}{\partial x} (x^0), \ldots, \frac{\partial L_f^{n-1} T_1 Z}{\partial x} (x^0)$ are linearly independent. On the other hand, by the chain of identities on the left column  of  \eqref{chain-identities-mimo}, we know that $L_f^i T_1 Z(x)= T_{i+1}Z(x)$, for $i=1,2,\ldots, n-1$, which implies that $\frac{\partial T_1 Z}{\partial x}(x^0), \frac{\partial T_2 Z }{\partial x}(x^0), \ldots, \frac{\partial T_n Z}{\partial x}(x^0)$ are linearly independent. Moreover, the right column of identities in  
\eqref{chain-identities-mimo} specialized to the case $m=1$ and 
evaluated at $x^0$ can be written as
\[
\begin{bmatrix}
\frac{\partial T_1 Z}{\partial x}(x^0)\\ 
\frac{\partial T_2 Z}{\partial x}(x^0)\\ 
\vdots\\ 
\frac{\partial T_{n-1} Z}{\partial x}(x^0)\\[1mm] 
\frac{\partial T_n Z}{\partial x}(x^0)
\end{bmatrix} g(x^0)=
\begin{bmatrix}
0\\ 
0\\ 
\vdots\\ 
0\\
M W(x^0)
\end{bmatrix} 
\]
This shows that $MW(x^0)\ne 0$, otherwise $g(x^0)=0$, which is a contradiction. 
Hence, $MW(x)\ne 0$ in a neighborhood of $x^0$ by continuity. 
\qedp

\subsection{Further details of Example \ref{ex:MIMO}}\label{subsec:details-example-MIMO}
The matrices $T,M,N$ associated to the other vectors of the basis are as follows
\[
\hspace{-0.2cm}
\begin{array}{l}
T^{(2)} = 
\left[
\begin{smallmatrix}
0_{3\times 1} &0_{3\times 1} &0_{3\times 1} &0_{3\times 1} &0_{3\times 1} &0_{3\times 1} &0_{3\times 1} &\\
0.0192 &  -0.1765 &  0 &   0.4965 &  -0.0192  &  0 &  0 &   0\\
0   & 0.0192 &  -0.1765  & 0  &  0.4965  &  0.1765 &  -0.1765  & 0
\end{smallmatrix}
\right] \\[4mm]
N^{(2)} = 
\left[
\begin{smallmatrix}
0_{3\times 1} & 0_{3\times 1} & 0_{3\times 1} & 0_{3\times 1} & 0_{3\times 1} & 0_{3\times 1} & 0_{3\times 1} & 0_{3\times 1}\\
0  &  0  &  0.0192 &  0 &  0 &  -0.0192 &   0.0192  &  0.4965
\end{smallmatrix}
\right]
\\[3mm]
M^{(2)}= 
\begin{bmatrix}
0 & 0 & 0 & 0\\
0 & 0.32 & -0.1765 & 0\\
\end{bmatrix}
\end{array}
\]

\[\begin{array}{l}
T^{(3)} = 
\left[
\begin{smallmatrix}
0_{3\times 1} &0_{3\times 1} &0_{3\times 1} &0_{3\times 1} &0_{3\times 1} &0_{3\times 1} &0_{3\times 1} &\\
0.4058 &  0.0455 &  0 &   -0.0117 &  -0.4058  &  0 &  0 &   0\\
0   & 0.4058 &  0.0455  & 0  &  -0.0117  &  -0.0455 &  0.0455  & 0
\end{smallmatrix}
\right] \\[4mm]
N^{(3)} = 
\left[
\begin{smallmatrix}
0_{3\times 1} & 0_{3\times 1} & 0_{3\times 1} & 0_{3\times 1} & 0_{3\times 1} & 0_{3\times 1} & 0_{3\times 1} & 0_{3\times 1}\\
0  &  0  &  0.4058 &  0 &  0 &  -0.4058 &   0.4058  &  -0.0117
\end{smallmatrix}
\right]
\\[3mm]
M^{(3)}= 
\begin{bmatrix}
0 & 0 & 0 & 0\\
0 & 0.0339 & 0.0455 & 0\\
\end{bmatrix}
\end{array}
\]

\[\begin{array}{l}
T^{(4)} = 
\left[
\begin{smallmatrix}
0_{3\times 1} &0_{3\times 1} &0_{3\times 1} &0_{3\times 1} &0_{3\times 1} &0_{3\times 1} &0_{3\times 1} &\\
0.0402 &  -0.3751 &  0 &   -0.1192 &  -0.0402  &  0 &  0 &   0\\
0   & 0.0402 &  -0.3751  & 0  &  -0.1192  &  0.3751 &  -0.3751  & 0
\end{smallmatrix}
\right] \\[4mm]
N^{(4)} = 
\left[
\begin{smallmatrix}
0_{3\times 1} & 0_{3\times 1} & 0_{3\times 1} & 0_{3\times 1} & 0_{3\times 1} & 0_{3\times 1} & 0_{3\times 1} & 0_{3\times 1}\\
0  &  0  &  0.0402 &  0 &  0 &  -0.0402 &   0.0402  &  -0.1192
\end{smallmatrix}
\right]
\\[3mm]
M^{(4)}= 
\begin{bmatrix}
0 & 0 & 0 & 0\\
0 & -0.4943 & -0.3751 & 0\\
\end{bmatrix}.
\end{array}
\]

\subsection{On the computation of the vector of basis functions $\phi(x,u)$ for Example \ref{ex:MIMO}}
Bearing in mind the matrices of functions $Z(x), Y(x), W(x)$, we compute 
\[
Z(x)\otimes  \frac{\partial Z}{\partial x}^\top=
\begin{bmatrix}
 x_1 & 0 & x_1 x_4 & 0 & 0\\
 0 & x_1 & 0 & 0 & 2 x_1 x_2\\
 0 & 0 & 0 & 0 & 0\\
 0 & 0 & x_1^2 & x_1 x_5 & 0\\
 0 & 0 & 0 & x_1 x_4 & 0\\
 x_2 & 0 & x_2 x_4 & 0 & 0\\
 0 & x_2 & 0 & 0 & 2 x_2^2\\
 0 & 0 & 0 & 0 & 0\\
 0 & 0 & x_1 x_2 & x_2 x_5 & 0\\
 0 & 0 & 0 & x_2 x_4 & 0\\
 x_1 x_4 & 0 & x_1 x_4^2 & 0 & 0\\
 0 & x_1 x_4 & 0 & 0 & 2 x_1 x_2 x_4\\
 0 & 0 & 0 & 0 & 0\\
 0 & 0 & x_1^2 x_4 & x_1 x_4 x_5 & 0\\
 0 & 0 & 0 & x_1 x_4^2 & 0\\
 x_4 x_5 & 0 & x_4^2 x_5 & 0 & 0\\
 0 & x_4 x_5 & 0 & 0 & 2 x_2 x_4 x_5\\
 0 & 0 & 0 & 0 & 0\\
 0 & 0 & x_1 x_4 x_5 & x_4 x_5^2 & 0\\
 0 & 0 & 0 & x_4^2 x_5 & 0\\
 x_2^2 & 0 & x_2^2 x_4 & 0 & 0\\
 0 & x_2^2 & 0 & 0 & 2 x_2^3\\
 0 & 0 & 0 & 0 & 0\\
 0 & 0 & x_1 x_2^2 & x_2^2 x_5 & 0\\
 0 & 0 & 0 & x_2^2 x_4 & 0
\end{bmatrix}
\]

\[
W(x)u= \begin{bmatrix}
u_1 & 0\\
0 & u_2 \\
u_1 c_{15}& 0\\
0 & u_2 c_{15} 
\end{bmatrix},\;
\frac{\partial Z}{\partial x}^\top
=
\begin{bmatrix}
 1& 0& x_4& 0& 0\\ 
 0& 1& 0& 0& 2 x_2\\ 
 0& 0& 0& 0& 0\\ 
 0& 0& x_1& x_5& 0\\ 
 0& 0& 0& x_4& 0
\end{bmatrix}
\]
from which we see that the functions in $W(x)u\otimes \frac{\partial Z}{\partial x}^\top$ are included in the set 
\[\ba{l}
\{
u, u_1 x, u_2 x, u_1 c_{15} x, u_2 c_{15} x
\}.
\ea
\]
Bearing in mind these expressions, the basis functions in $\phi(x,u)$ can be chosen as  
\[\ba{l}
x, x_1^2, x_2^2, x_1 x_2, x_1 x_4, x_1 x_5, x_2 x_4, x_2 x_5, x_4 x_5, x_1 x_2 x_4, \\
  x_1 x_4 x_5, x_1^2 x_4, x_1 x_4^2, x_2^3, x_2^2 x_4, x_2^2 x_5, x_2 x_4 x_5, x_4 x_5^2, x_4^2 x_5,\\ 
  u, u c_{15}, u_1 x, u_2 x, u_1 c_{15} x, u_2 c_{15} x.
\ea
\]

\bibliographystyle{IEEEtran}

\bibliography{refs}


%
%
%
%

\end{document}